\documentclass[a4paper,12pt]{article}
\usepackage{amsfonts,amsmath,amssymb,color,comment,listings}
\usepackage[paper=letterpaper,margin=1.0in]{geometry}
\usepackage{graphicx,ytableau}
\usepackage{cite}
\usepackage{hyperref}
\usepackage{color}
\usepackage{xcolor}
\usepackage{young} 
\usepackage{tikz,pgfplots}
\usetikzlibrary{angles,quotes,arrows.meta,chains,matrix,scopes,calc,intersections,positioning,shapes.misc,decorations.markings,shapes.geometric}
\tikzset{cross/.style={cross out, draw=black,thick, minimum size=3*(#1-\pgflinewidth), inner sep=0pt, outer sep=0pt},
cross/.default={3pt}}

\newcommand{\be}{\begin{equation}}
\newcommand{\ee}{\end{equation}}
\newcommand{\bea}{\begin{eqnarray}\displaystyle}
\newcommand{\eea}{\end{eqnarray}}

\newcommand{\ba}{\begin{array}}
\newcommand{\ea}{\end{array}}
\newcommand{\ben}{\begin{enumerate}}
\newcommand{\een}{\end{enumerate}}
\newcommand{\bi}{\begin{itemize}}
\newcommand{\ei}{\end{itemize}}
\newcommand{\bc}{\begin{center}}
\newcommand{\ec}{\end{center}}
\newcommand{\bfig}{\begin{figure}}
\newcommand{\efig}{\end{figure}}
\newcommand{\bq}{\begin{quotation}}
\newcommand{\eq}{\end{quotation}}
\newcommand{\bt}{\begin{table}}
\newcommand{\et}{\end{table}}
\newcommand{\btab}{\begin{tabular}}
\newcommand{\etab}{\end{tabular}}
\newcommand{\bmi}{\begin{minipage}}
\newcommand{\emi}{\end{minipage}}
\newcommand{\bs}{\begin{slide}}
\newcommand{\es}{\end{slide}}

\newcommand{\wt}{\widetilde}

\def\wt{ \widetilde }

\def\c2{ \widehat \chi_2^R }

\newtheorem{definition}{Definition}
\newtheorem{conjecture}{Conjecture}
\newtheorem{corollary}{Corollary}

\title{\bf{\huge{A generalized dominance ordering for 1/2-BPS states}}}
\author{\bf{\Large{Garreth Kemp$^{1}$\footnote{garry@kemp.za.org}}}}
\date{}			

\begin{document}

\begin{titlepage}
\maketitle

\begin{center}
	\emph{$^{1}$Department of Mathematics and Applied Mathematics, \\
	University of Johannesburg, Auckland Park, 2006,\\
	South Africa}
\end{center}

\begin{abstract}

We discuss a generalized dominance ordering for irreducible representations of the symmetric group $S_{n}$ with the aim of distinguishing the corresponding states in the 1/2-BPS sector of $U(N)$ Super Yang-Mills theory when a certain finite number of Casimir operators are known. Having knowledge of a restricted set of Casimir operators was proposed as a mechanism for information loss in this sector and its dual gravity theory in AdS$_{5}\times S^{5}$. It is well-known that the states in this sector are labeled by Young diagrams with $n$ boxes. We propose a generalization of the well-known dominance ordering of Young diagrams. Using this generalization, we posit a conjecture to determine an upper bound for the number of Casimir operators needed to distinguish between the 1/2-BPS states and thus also between their duals in the gravity theory. We offer numerical and analytic evidence for the conjecture. Lastly, we discuss implications of this conjecture when the energy $n$ of the states is asymptotically large.

\end{abstract}
\end{titlepage}


\tableofcontents

\section{Introduction}

 The 1/2-BPS sector of $\mathcal{N}=4$ SYM with $U(N)$ gauge group has a gravitational dual in AdS$_{5}\times S^{5}$ which contains many interesting states including giant gravitons \cite{GGinCFT} and LLM geometries \cite{LLM}. When the energy of 1/2-BPS states scale like $O(N)$, they are dual to giant gravitons and when they scale as $O(N^{2})$, they are dual to LLM geometries. Operators in the CFT dual to these states were constructed and are organised by irreducible representations $R$ of $U(N)$ \cite{CJR}. These operators constitute an exactly orthogonal gauge invariant basis for this sector. These are known as the Schur polynomials and are labeled by Young diagrams $R$ with $n$ boxes, where $n$ is the energy of the corresponding state. These operators, as well as group representation theory, have played an important role in the physics of correlators involving giant gravitons and the LLM geometries - see for example \cite{Hai}, \cite{BKY}, \cite{DmK}. It is known that having complete knowledge of all $N$ Casimir operators in the $U(N)$ gauge theory allows one to uniquely identify an irreducible representation \cite{Zelo}. These Casimir operators are related to conserved charges \cite{CJR} and it was argued in \cite{BCLS} that all $N$ conserved charges are encoded in the multipole moments in the dual gravitational solution for the LLM geometries. However, they further argued that even a semi-classical observer performing measurements of the multipole moments up to the Planck scale limit $O(N^{1/4})$ will not acquire the information needed to reconstruct or identify the state. Consequently, information is necessarily lost for a semi-classical observer attempting to uniquely identify the irrep labelling the state. See also \cite{BC} and \cite{Beren} for further interesting discussions on this topic.

This inspired the work of \cite{KempRam} in which information loss was studied and quantified when a restricted set of Casimir operators is known  and the states have a fixed energy $n$. The Casimir operators $C_{k}$ are related, through Schur-Weyl duality, to elements in the centre of the Symmetric group algebra $\mathbb{C}(S_{n})$ called cycle operators, $T_{k}$. These central elements are defined to be sums of permutations of $S_{n}$ with the cycle structure consisting of a single cycle of length $k$, and all other cycles of length 1. In \cite{KempRam}, structural properties of the group algebra $\mathbb{C}(S_{n})$ and its centre $\mathcal{Z}[\mathbb{C}(S_{n})]$ were studied. For example, the set of cycle operators $\{T_{2}, \cdots , T_{n}\}$ is capable of multiplicatively generating $\mathcal{Z}[\mathbb{C}(S_{n})]$. This is related to the list of normalized characters $\{ \widehat{\chi}_{R}(T_{2}) , \widehat{\chi}_{R}(T_{3}) , \cdots , \widehat{\chi}_{R}(T_{n})  \}$ being distinct amongst the irreps of $S_{n}$ labeled by Young diagrams $R$ of $n$ boxes. The list of normalized characters, being unique for each $R$, is thus capable of distinguishing all the $S_{n}$ irreps. Interestingly, only a subset $\{ T_{2} , \cdots , T_{k} \}$ suffices to generate the centre, where $k < n$. The subset is capable of generating $\mathcal{Z}[\mathbb{C}(S_{n})]$ if and only if the list $\{ \widehat{\chi}_{R}(T_{2}) , \cdots , \widehat{\chi}_{R}(T_{k}) \}$ is unique over the irreps $R$, in which case distinguishing between the irreps is possible. As $n$ increases, with $k$ held fixed, the subset of central elements fails to generate the centre, and instead generates only a subspace of the centre. Related to this is the fact that degeneracies appear amongst the lists of normalized characters. Once degeneracies appear, one then extends the subset to include $T_{k+1}$. One examines the list of normalized characters of $\{T_{2} , \cdots , T_{k} ,  T_{k+1}\}$ for uniqueness amongst the irreps, $R$. The integer $k^{(n)}_{*}$ is defined as the smallest number of $T_{k}$ capable of generating $\mathcal{Z}[\mathbb{C}(S_{n})]$. 

The structural properties of $\mathbb{C}(S_{n})$ and its centre, were used to  define solvable quantum mechanics models \cite{GelounRam0}. The cycle operators were used to construct Hamiltonians whose eigenvalues were again the normalized characters and multiplicities were the representation theoretic objects known as the Kronecker coefficients. Furthermore, this work provided algorithms for combinatoric calculation of the Kronecker coefficients. Using these quantum mechanical models, connections were made with quantum computation and complexity theory. The complexity of these algorithms depended crucially on the integers $k^{(n)}_{*}$.

The task of distinguishing between the irreps, or Young diagrams, of $S_{n}$ with a restricted number of cycle operators $\{T_{2} , \cdots , T_{k}\}$ raises a number of interesting questions. For example, what are the properties of the Young diagrams whose normalized characters are degenerate? Through Schur-Weyl duality this question translates to properties of 1/2-BPS states in the $U(N)$ gauge theory, and through AdS/CFT, to properties of the dual geometries in AdS. Next, what is the asymptotic behaviour of $k^{(n)}_{*}$ when $n$ is large? An elegant heuristic argument in \cite{GelounRam1} finds that $k^{(n)}_{*} \sim n^{1/4}$ for large $n$. A quantum phase estimation algorithm \cite{Nielsenandchuang} is given for detecting Young diagrams, which also label a combinatoric basis for $\mathcal{Z}[\mathbb{C}(S_{n})]$, with $k^{(n)}_{*}$ again playing a pivotal role in quantifying the complexity of the algorithm. The results in \cite{GelounRam1} established once again deep connections between the structural properties of $\mathbb{C}(S_{n})$, and its centre, with quantum computational and complexity theory. Obtaining a deeper understanding of $k^{(n)}_{*}$ can be expected to have interesting implications, not only in holography, but also quantum computation and quantum complexity. 

In this work, we provide a conjecture that will allow us to establish an upper bound for $k_{*}^{(n)}$ based on the structure of the Young diagrams labelling the $1/2$-BPS states. The structure of the Young diagrams we are concerned with here is based on the majorization, or dominance, ordering of partitions of $n$ which we define in section \ref{domincomp}. We find that only when two Young diagrams $Y$ and $Y'$, both with $n$ boxes, are incomparable according to the dominance ordering scheme can they have degenerate normalized characters for the cycle operators starting from $T_{2}$, progressing to $\{T_{2} , T_{3}\}$ and so on. We introduce an integer quantifying the incomparability between the pair $(Y,Y')$, and conjecture that the size of the subset $\{T_{2} , T_{3} , \cdots , T_{k} \}$ having degenerate normalized characters for $(Y,Y')$ is bounded by our incomparability measure. We also provide numerical and analytical evidence in support of the conjecture. The theory of majorization has found interesting applications in recent years in the field of quantum information relating to entanglement transformation using local operations and classical communication (LOCC) \cite{NielsenEnt}, \cite{JonathanPlenio}. A quantum state $\rho$ with a given level of entanglement may be transformed into another state $\sigma$ with a different amount of entanglement through LOCC if and only if the vector of eigenvalues of $\rho$ majorizes the vector of eigenvalues of $\sigma$. Incomparable partitions have also played an interesting role in this story. If the two states are incomparable then entanglement transformation using LOCC may still be possible by using third state called a catalyst \cite{Cat0}, \cite{Cat01}, \cite{Cat1} and \cite{Cat2}. A proof is given in \cite{Turgrut} for necessary and sufficient conditions for when this catalytic transformation is possible. Power sum symmetric polynomials (or power sums for short) evaluated on the partitions play a central role in this proof. It may be interesting to note that the power sums appearing in section \ref{contpolypowersums} of this work are reminiscent of those appearing in \cite{Turgrut}. Lastly, incomparable partitions have also have found interesting applications in statistical physics \cite{Boltzman1} \cite{Boltzman2} discussing so-called mixedness and complexity of Boltzmann macrostates.

This paper is structured as follows. In section \ref{domincomp} we outline some background on the dominance ordering we need later on. We also define an integer quantifying, or measuring, the incomparability between two Young diagrams with $n$ boxes. In section \ref{conjecture} we state our conjecture for a necessary condition for normalised characters of the subset $\{T_{2} , \cdots , T_{k}\}$ to be degenerate in irreps of $S_{n}$. This culminates in a conjectured upper bound for $k^{(n)}_{*}$. In section \ref{Data} we show some data generated in Mathematica supporting the conjecture. In section \ref{contpolypowersums} we use known formulas in the mathematics literature to express these normalized characters in terms of content polynomials. Thereafter, we derive expressions for the content polynomials in terms of power sums in the shifted row lengths of the corresponding Young diagrams. We then restate our conjecture for the degenerate normalized characters in terms of these power sums. In section \ref{analarguments} we make use of the conditions and formulas derived in section \ref{contpolypowersums} to present some analytic evidence for our conjecture.  Section \ref{Asymp} discusses an algorithm for constructing a pair of Young diagrams having a given incomparability (according to the measure we define) using the least number of boxes, $n$. From this, we study the asymptotic behaviour of the $k^{(n)}_{*}$ upper bound. We wrap up the paper with a discussion. There are four appendices. Appendices \ref{2swap}, \ref{DeltaID} and \ref{Transitivity} contain some explicit calculations and additional content pertaining to section \ref{analarguments}. Appendix \ref{ConstructPart} contains examples of the material discussed in section \ref{Asymp}.

\section{Majorization and incomparable partitions of $n$}\label{domincomp}

The concept of majorization is a way to compare two partitions, or a set of partitions. Majorization for partitions is defined in the following way. Consider two partitions of $n$, $X = ( x_{1} , x_{2} , \cdots , x_{d} )$ and $ Y = ( y_{1} , y_{2} , \cdots , y_{d} )$ with $x_{1} \geq x_{2} \geq \cdots \geq x_{d}$ and $\sum^{d}_{i=1}x_{i} = n$, and similarly for $Y$. $Y$ is said to majorize $X$, written $X \prec Y$, if
\bea
\label{eq:1a}
	\sum\limits^{K}_{i=1} x_{i} \leq \sum\limits^{K}_{i=1} y_{i} , \hspace{20pt} K = 1,2, \cdots, d-1,
\eea
with equality when $K=d$. Of course, for $X\neq Y$, then some of the inequality signs are strict inequalities. Generically, the majorization definition holds for vectors $v$ and $w$ whose components need not have any particular order. However, when restricting to partitions whose components are ordered in a weakly decreasing fashion, majorization, also sometimes called the dominance ordering, establishes a partial ordering on the space of partitions. See \cite{MOA} for more discussions with many applications. The ordering is partial since it is possible for two partitions to be incomparable according to this scheme. For example, the set of partitions of 6 contains $(4,2,0)$ and $(3,2,1)$. Studying the row sum relations in (\ref{eq:1a}), we see that $(3,2,1) \prec (4,2,0)$. However, the partitions $(4,1,1)$ and $(3,3,0)$ are incomparable since, looking at the row sums,
\bea
	3 &<& 4,\\ \nonumber
	6 &>& 5, \\
	6 &=& 6.\nonumber
\eea
At $n=6$, there are only two pairs of partitions that are incomparable. These are $\{ (4,1,1), (3,3,0) \}$ and their transposes $\{ ( 3,1,1,1 ) , (2,2,2,0) \}$. At higher $n$, there exist partitions such that, for a given incomparable pair, the inequality sign in their row sum relations swaps more than once. As an illustration, $(8, 6, 6, 5, 1, 1, 1)$ and $(7, 7, 7, 3, 2, 2)$ are both partitions of 28 and have three swaps:
\bea
	7 &<& 8, \nonumber \\
	14 &=& 14, \nonumber \\
	21 &>& 20,\nonumber\\
\label{eq:1b}
	24&<& 25,  \\
	26&=& 26, \nonumber\\
	28&>& 27,\nonumber\\
	28&=& 28.\nonumber
\eea
In the above row sums, $(8, 6, 6, 5, 1, 1, 1)$ starts off dominating $(7, 7, 7, 3, 2, 2)$ but then the inequality sign in the row sums subsequently swaps direction three times.  
\begin{definition}
	Let $X$ and $Y$ be two partitions of $n$ of the same length $d$. The symbol of $X$ and $Y$ is defined as the $d$-tuple $(s_{1}, s_{2},\cdots, s_{d})$, where $s_{i} \in \{<, >, =\}$, recording the row sum relations between the pair $(X, Y)$.
\end{definition}
For example, in (\ref{eq:1b}) the symbol for the two partitions is $(<,=,>,<,=,>,=)$.

We introduce a measure of the incomparability of two Young diagrams $X$ and $Y$ with an integer $k^{(X,Y)}_{swap}$. 
\begin{definition}
	For partitions $X$ and $Y$ of n, define $k^{(X,Y)}_{swap}$ to be an integer counting the number of swaps in the direction the inequality sign faces in the row sum relations of $X$ and $Y$.
\end{definition}
For the previous example, $k_{swap} = 3$. The higher $k^{(X,Y)}_{swap}$, the more incomparable $X$ and $Y$ are. If $k^{(X,Y)}_{swap} = 0$, then $X$ and $Y$ are comparable, since either $X\prec Y$ or $Y\prec X$. This integer plays a central role in our conjecture stated in section \ref{conjecture}.

In what follows below, partitions with shifted row lengths will also be important. Let $Y$ be a partition of $n$. Then we define $\widetilde{Y}$ to be
\bea
	\widetilde{Y} = (y_{1} - 1 , y_{2} - 2 , \cdots , y_{d} - d ).
\eea
Note that the majorization relations, or the dominance ordering, between $\widetilde{X}$ and $\widetilde{Y}$ will be the same as for $X$ and $Y$. If $X\prec Y$ , then we still have $\widetilde{X} \prec \widetilde{Y}$. If $X$ and $Y$ are incomparable with $k^{(X,Y)}_{swap} = l$, then for the corresponding $\widetilde{X}$ and $\widetilde{Y}$ we also have $k^{(\widetilde{X},\widetilde{Y})}_{swap} = l$. 

\section{Statement of the conjecture}\label{conjecture}

We begin this section by defining two integers:
\begin{definition}\label{Def3}
	 Define the integer $k^{(Y_{1} , Y_{2})}_{*}$ for Young diagrams $Y_{1}$ and $Y_{2}$ of $n$ boxes to be the minimum number of cycle operators $T_{k}$ needed to distinguish $Y_{1}$ from $Y_{2}$.
\end{definition} 

\begin{definition}
	Define the integer $k^{(n)}_{*}$ to be the minimum number of cycle operators $T_{k}$ needed to distinguish all irreps at a given $n$:
	\bea
		k^{(n)}_{*} = \underset{X,Y \vdash n }{\max} k^{(X,Y)}_{*}.
	\eea
\end{definition}

Consider a pair of Young diagrams $(Y_{1} , Y_{2})$, both partitions of $n$. Different pairs may have different degeneracy in the list of normalized characters. For example, the pair $(Y_{1} , Y_{2})$ may have degenerate normalized characters for $T_{2}$ only. But $(Y_{2} , Y_{3})$ may have degenerate $\{T_{2},T_{3}\}$\footnote{Of course, this means that the pair $(Y_{1} , Y_{3})$ will also have degenerate $T_{2}$.}. From Definition \ref{Def3}, the normalized characters of $\{ T_{2} , \cdots , T_{k^{(Y_{1} , Y_{2})}_{*}}\}$ will distinguish $Y_{1}$ from $Y_{2}$. Let $Y_{1}$ and $Y_{2}$ have degenerate normalized characters for $T_{2}$ up to some  $T_{k^{(Y_{1},Y_{2})}_{deg}} $, where $k^{(Y_{1},Y_{2})}_{deg}$ is some integer larger than 1. Thus,
\begin{eqnarray}
	\widehat{\chi}_{Y_{1}}(T_{2}) &=& \widehat{\chi}_{Y_{2}}(T_{2})  \nonumber \\
	\widehat{\chi}_{Y_{1}}(T_{3}) &=& \widehat{\chi}_{Y_{2}}(T_{3})  \nonumber \\
\label{eq:2a}
	&\vdots&\\
	\widehat{\chi}_{Y_{1}}(T_{k^{(Y_{1},Y_{2})}_{deg}}) &=& \widehat{\chi}_{Y_{2}}(T_{k^{(Y_{1},Y_{2})}_{deg}}) , \nonumber 
\end{eqnarray}
with $\hat{\chi}\big(T_{k^{(Y_{1},Y_{2})}_{deg}+1}\big)$ distinct for $Y_{1}$ and $Y_{2}$. Then $k^{(Y_{1} , Y_{2})}_{*} = k^{(Y_{1},Y_{2})}_{deg} + 1$.\\ 
\begin{conjecture}
 If the normalized characters of $\{T_{2} , T_{3} , \cdots , T_{k^{(Y_{1},Y_{2})}_{deg}} \}$ are degenerate for the pair $(Y_{1} , Y_{2})$, then
\begin{eqnarray}
\label{eq:2b}
	 k^{(Y_{1},Y_{2})}_{deg}  \leq k^{(Y_{1} , Y_{2})}_{swap} + 1. 
\end{eqnarray}
\end{conjecture}
A simple corollary follows:
\begin{corollary} 
An upper bound on $k^{(Y_{1} , Y_{2})}_{*} $ can be given in terms of $ k^{(Y_{1} , Y_{2})}_{swap}$:
\begin{eqnarray}
\label{eq:2c}
	k^{(Y_{1} , Y_{2})}_{*} \leq k^{(Y_{1} , Y_{2})}_{swap} + 2 .
\end{eqnarray}
\end{corollary}
Thus, for $Y_{1}$ and $Y_{2}$, the number of cycle operators $\{T_{2} , \cdots , T_{k^{(Y_{1},Y_{2})}_{deg}}\}$ whose normalized characters are degenerate is bounded above by how incomparable $Y_{1}$ and $Y_{2}$ are as measured by $k^{(Y_{1} , Y_{2})}_{swap}$. It is also interesting to ask if one can similarly express a lower bound for $k^{(Y_{1},Y_{2})}_{*}$ in terms of characteristics of Young diagram pairs. We do not have anything to add in this direction at this time. 
 
Let us explore some implications of this conjecture and corollary. Given any two partitions $(Y_{1}, Y_{2})$ of $n$ we can easily ascertain $k^{(Y_{1} , Y_{2})}_{swap}$ simply by inspection. Once we know $k^{(Y_{1} , Y_{2})}_{swap}$ we have an upper bound on $k^{(Y_{1},Y_{2})}_{deg}$, and an upper bound on $k^{(Y_{1} , Y_{2})}_{*}$. 

Say, we have a pair $Y_{1}$ and $Y_{2}$ for which $k^{(Y_{1} , Y_{2})}_{swap} = 2$. Then according to the conjecture, $k^{(Y_{1},Y_{2})}_{deg} \leq 3$ and according to the corollary $k^{(Y_{1} , Y_{2})}_{*} \leq 4$. We conclude that the list of normalized characters of $\{T_{2},T_{3} ,T_{4}\}$ is guaranteed to distinguish $Y_{1}$ and $Y_{2}$. Now, it is possible for $k^{(Y_{1},Y_{2})}_{deg} = 1$, which means that $T_{2}$ is distinct for $Y_{1}$ and $Y_{2}$. Thus, $T_{2}$ is able to distinguish $Y_{1}$ from $Y_{2}$ and $k^{(Y_{1},Y_{2})}_{*} = 2$. It is also possible for $k^{(Y_{1} , Y_{2})}_{deg} = 2$, i.e. only $T_{2}$ is degenerate. Then we need only $\{T_{2} , T_{3}\} $ to distinguish the pair, since $T_{3}$ is distinct. So $k^{(Y_{1} , Y_{2})}_{*} = 3$. On the other hand, it is also possible for $\{ T_{2} , T_{3} \} $ could be degenerate, making $k^{(Y_{1},Y_{2})}_{deg} = 3$, and $k^{(Y_{1},Y_{2})}_{*} = 4$. According to (\ref{eq:2b}) , this is the maximum possible value for $k^{(Y_{1} , Y_{2})}_{deg}$. From (\ref{eq:2c}), we see that  $k^{(Y_{1} , Y_{2})}_{*} = 4$ has saturated the bound. Thus, we do not need more than $\{ T_{2} , T_{3} , T_{4} \}$ to distinguish $Y_{1}$ from $Y_{2}$. 

Suppose we know the maximum number of swaps any pair of partitions can have at some fixed $n$. Call this $k^{(n)}_{(swap,max)}$. 
\begin{definition}
	Define $k^{(n)}_{(swap,max)}$ to be the maximum number of swaps of the inequality sign in the row sum relations for all pairs of Young diagrams of $n$ boxes.
	\bea
	\label{eq:2ca}
		k^{(n)}_{(swap,max)} = \underset{X,Y \vdash n }{\max} k^{(X,Y)}_{swap}.
	\eea
\end{definition}
Then we have a bound for $k^{(n)}_{*}$:
\begin{eqnarray}
\label{eq:2d}
	k^{(n)}_{*} \leq k^{(n)}_{(swap,max)} + 2 .
\end{eqnarray}
Thus, we conclude that the normalized characters of $\{T_{2} , \cdots , T_{k^{(n)}_{(swap,max)} + 2} \}$ are guaranteed to distinguish all the irreps at this $n$.

To illustrate these ideas, consider a specific example, say $n = 27$\footnote{There is no particular reason for studying $n=27$ specifically - the number of partitions is large enough to provide a rich enough example to clearly illustrate the implications of (\ref{eq:2d}). }. There are a total of 3010 partitions of $27$. Conjecture \ref{Conj2} in section \ref{Asymp}, and the subsequent data in Table \ref{Tab:Tabnmin}, shows that the maximum number of swaps any pair of these partitions can have is 4.
 \begin{itemize}
 \item At this $n$ value, there exist pairs $(Y_{1} , Y_{2})$ for which $k^{(Y_{1} , Y_{2})}_{swap} = i$ with $i=0,1,2,3.$ From our conjecture (\ref{eq:2c}) $k^{(Y_{1},Y_{2})}_{deg} \leq i+1$ respectively thus making $k^{(Y_{1} , Y_{2})}_{*} \leq i+2$ respectively for these pairs. Of course, if the bound for $k^{(Y_{1},Y_{2})}_{deg}$ is saturated ($k^{(Y_{1} , Y_{2})}_{deg} = i+1$), then the bound for $k^{(Y_{1} , Y_{2})}_{*}$ is also saturated ($k^{(Y_{1} , Y_{2})}_{*} = i+2$).
  \item Finally, we could have pairs for which $k^{(Y_{1} , Y_{2})}_{swap} = k^{(n=27)}_{(swap,max)} = 4$. From the conjecture, $k^{(Y_{1} , Y_{2})}_{deg} \leq 5$, implying that $k^{(Y_{1} , Y_{2})}_{*} \leq 6$ for all such pairs. Thus, the set $\{T_{2} , T_{3} , T_{4}, T_{5}, T_{6} \}$ is guaranteed to distinguish all partition pairs having 4 swaps in their row sum relations. But also, the set is guaranteed to distinguish all other pairs since all other pairs have fewer swaps in their row sums. Thus $k^{(n)}_{*} \leq 6$ for $n=27$. Again, if the bound for $k^{(Y_{1} , Y_{2})}_{deg}$ is saturated then the bound for $k^{(Y_{1} , Y_{2})}_{*}$ will also be saturated. Using known formulae for calculating normalized characters \cite{Lassalle} , \cite{CGS}, we have established that $k^{(n)}_{*} = 5$ at $n = 27$ \cite{KempRam}.
\end{itemize} 

To summarize, we have stated a conjecture, based on the properties of partitions of $n$, that allows us to place an upper bound on the integer $k^{(n)}_{*}$. Counting the number of times the inequality sign switches direction in the row sum relations is playing a key role in this conjecture. The integer $k^{(Y,Y')}_{swap}$ is a measure of how incomparable partitions $Y$ and $Y'$ are. We emphasize that $k^{(Y,Y')}_{swap} > 0$ is a necessary but not sufficient condition for $Y$ and $Y'$ to have degenerate normalized characters of cycle operators starting from $T_{2}$ proceeding to $T_{3}$ and so on.

\section{Data supporting the conjecture}\label{Data}

In this section we present data in which we count the number of times the inequality sign swaps direction in the row sum relations for partition pairs $Y$ and $Y'$ having a given degeneracy in the list of normalized characters. Concretely, we consider all Young diagrams of $n$ boxes having degenerate $(\widehat{\chi}_{Y}(T_{2}) , \cdots , \widehat{\chi}_{Y}(T_{k_{deg}}) )$ for a fixed $k_{deg}$. This will give subsets of partitions having this degeneracy with the actual values of the list of normalized characters differing for each subset. In each subset, we pair the Young diagrams in all possible ways, and we measure $k^{(Y,Y')}_{swap}$ for each of these pairs. We then repeat the procedure for the next subset of partitions having degenerate $(\widehat{\chi}_{Y}(T_{2}) , \cdots , \widehat{\chi}_{Y}(T_{k_{deg}}) )$. Finally, we tally the total number of pairs having a particular number of swaps. Once we have this data, we compare with equation (\ref{eq:2b}) to test our conjecture. Importantly, for our conjecture to be satisfied, we need to detect zero Young diagrams pairs having fewer than $k_{deg} - 1$ swaps for a given fixed $k_{deg}$. We then move on to the next $n$ value, repeating the procedure described above.

As an illustrative example, for Young diagrams at a given $n$, consider the case of $k_{deg} = 3$. Say the Young diagrams in each of the following subsets,
   \begin{eqnarray}
      \label{eq:4a}
	 \{ Y_{1} , Y_{2} \} , \hspace{10pt} \{ Y_{3} , Y_{4} , Y_{5} \} ,\hspace{10pt} \{ Y_{6} , Y_{7}  \} . 
\end{eqnarray}  
have degenerate values for $(\widehat{\chi}_{Y}(T_{2}) , \widehat{\chi}_{Y}(T_{3}) )$. For clarity, note that the values of these normalised characters will differ in general from subset to subset. Now we pair the diagrams in each subset in all possible ways, and measure $k_{swap}$ for each pair. This means we obtain 
   \begin{eqnarray}
      \label{eq:4b}
	\{k^{(Y_{1},Y_{2})}_{swap} \}, \hspace{10pt} \{k^{(Y_{3},Y_{4})}_{swap}, k^{(Y_{3},Y_{5})}_{swap},k^{(Y_{4},Y_{5})}_{swap}\} , \hspace{10pt}  \{k^{(Y_{6},Y_{7})}_{swap} \}. 
\end{eqnarray}
Say we find for the above $k_{swap}$ values (ignoring our conjecture for the moment):
\begin{eqnarray}
      \label{eq:4c}
	\{ 2 \}, \hspace{10pt} \{ 1 , 0, 2 \} , \hspace{10pt}  \{ 1 \}. 
\end{eqnarray}  
 We then tally up the $k_{swap}$ values:
\begin{itemize}
  \item We have 1 pair with 0 swaps,
  \item We have 2 pairs with 1 swap,
  \item We have 2 pairs with 2 swaps.
\end{itemize}
This hypothetical example for $k_{deg} = 3$ is not consistent with the conjecture. 

The data we present begins from the smallest $n$ value for which Young diagram pairs appear having a particular $k_{deg}$. This was studied for small values of $k_{deg}$ in \cite{KempRam}. This information is summarized in Table \ref{Tab:Tabnminforkdeg}. For example, the first instance of partitions appearing with $k_{deg} = 2$ is for $n=6$. Thus, for $k_{deg} = 2$ we begin our data from $n = 6$.

\begin{table}[htp]
\begin{center}
\begin{tabular}{|c|c|}
\hline
 $k_{deg}$ & first $n$  \\
 \hline
 2 & 6 \\
 \hline
 3 & 15 \\
 \hline
 4 & 24 \\
  \hline
 5 & 42 \\
  \hline
 6 & 80 \\
  \hline
\end{tabular}
\caption{\small{Table showing the first $n$ value at which Young diagrams appear having a particular degeneracy $k_{deg}$. See \cite{KempRam} for the details }}
\label{Tab:Tabnminforkdeg}
\end{center}
\end{table}
Our data is presented in Tables \ref{Tab:T2data}, \ref{Tab:T2T3data}, \ref{Tab:T2T3T4data}, \ref{Tab:T2T3T4T5data}. Table \ref{Tab:T2data} considers $k_{deg} =2$ from $n=6$ to $n=27$, Table \ref{Tab:T2T3data} conisders $k_{deg} = 3$ from $n=15$ to $n=42$, Table \ref{Tab:T2T3T4data} considers $k_{deg} = 4$ from $n=24$ to $n=48$, and the last table, \ref{Tab:T2T3T4T5data}, is for $k_{deg} = 5$ from $n = 42$ to $n=65$. Importantly, in order to support the conjecture, we should detect zero Young diagram pairs having fewer than $k_{deg} - 1$ swaps. 
\begin{table}{}
  \centering 
  \begin{tabular}{|c|c|c|c|c|c|}
\hline
 $n$  & 0 swaps & 1 swap  & 2 swaps & 3 swaps & 4 swaps  \\
 \hline
  6 & 0 & 2  & 0  & 0  & 0 \\
  \hline
  7 & 0 & 0 &0  &0  & 0 \\
  \hline
  8 & 0 & 3 & 0 & 0 & 0\\
  \hline
  9 & 0 & 5 & 0 & 0 & 0\\
  \hline
  10 & 0 & 15 &0 & 0 & 0\\
  \hline
  11 & 0 & 9 & 0& 0 & 0\\
  \hline
  12 & 0 & 49 &0 & 0 &0 \\
  \hline
  13 & 0 & 35 &0 & 0 & 0\\
  \hline
  14 & 0 & 91 & 0& 0 & 0\\
  \hline
  15 & 0 & 156 & 3 & 0 &0 \\
  \hline
  16 & 0 & 258 & 6 & 0 & 0\\
  \hline
  17 & 0 & 290 & 8 & 0 & 0\\
  \hline
  18 & 0 & 644 & 32 & 2 &0 \\
  \hline
  19 & 0 & 699 & 19 & 0 &0 \\
  \hline
  20 & 0 & 1371 & 81 & 4  &0 \\
  \hline
  21 & 0 & 1905 &  107 &7  &0 \\
  \hline
  22 & 0 & 2974 & 215  &  12 &0 \\
  \hline
  23 & 0 & 3593 & 230 & 15 & 0\\
  \hline
  24 & 0 & 6892 & 637 & 51 &0 \\
  \hline
  25 & 0 & 8343 & 704 & 55 &0 \\
  \hline
  26 & 0 & 13273 & 1329 & 127 &0 \\
  \hline
  27 & 0 & 18099 & 1927 & 188 &0 \\
  \hline
\end{tabular}
  \caption{This table measures $k_{swap}$ for all pairs of diagrams having degenerate $\widehat{\chi}_{Y}(T_{2})$. We only considered up to four swaps. We detect zero Young diagram pairs having degenerate $\{ \widehat{\chi}_{Y}(T_{2})  \}$ having less than one swap.}
  \label{Tab:T2data}
\end{table}
\begin{table}{}
  \centering 
  \begin{tabular}{|c|c|c|c|c|c|}
\hline
 $n$  & 0 swaps & 1 swap  & 2 swaps & 3 swaps & 4 swaps  \\
 \hline
  15 & 0 & 0  & 3  & 0  & 0 \\
  \hline
  16 & 0 & 0 & 4  &0  & 0 \\
  \hline
  17 & 0 & 0 & 4 & 0 & 0\\
  \hline
  18 & 0 & 0 & 8 & 0 &0 \\
  \hline
  19 & 0 & 0 & 1 & 0 &0 \\
  \hline
  20 & 0 & 0 & 17 & 0  &0 \\
  \hline
  21 & 0 & 0 &  21 & 0  &0 \\
  \hline
  22 & 0 & 0 & 31  & 0 &0 \\
  \hline
  23 & 0 & 0 & 36 & 0 & 0\\
  \hline
  24 & 0 & 0 & 93 & 1 &0 \\
  \hline
  25 & 0 & 0 & 104 & 0 & 0 \\
  \hline
  26 & 0 & 0 & 125 & 0 &0 \\
  \hline
  27 & 0 & 0 & 181 & 2 &0 \\
  \hline
  28 & 0 & 0 & 281 & 2 &0 \\
  \hline
  29 & 0 & 0 & 304 & 0 &0 \\
  \hline
  30 & 0 & 0 & 563 & 6 & 1 \\
  \hline
  31 & 0 & 0 & 616 & 5 & 2 \\
  \hline
  32 & 0 & 0 & 975 & 22 &0 \\
  \hline
  33 & 0 & 0 & 1209 & 25 & 7 \\
  \hline
  34 & 0 & 0 & 1537 & 30 & 4 \\
  \hline
  35 & 0 & 0 & 2167 & 65 & 8 \\
  \hline
  36 & 0 & 0 & 3154 & 110 & 15 \\
  \hline
  37 & 0 & 0 & 3748 & 114 & 21 \\
  \hline
  38 & 0 & 0 & 4873 & 150 & 18 \\
  \hline
  39 & 0 & 0 & 6401 & 223 & 28 \\
  \hline
  40 & 0 & 0 & 9101 & 425 & 70 \\
  \hline
  41 & 0 & 0 & 10909 & 389 & 65 \\
  \hline
  42 & 0 & 0 & 15045 & 714 & 147 \\
   \hline
\end{tabular}
  \caption{This table measures $k_{swap}$ for all pairs of diagrams having degenerate $\left(\widehat{\chi}_{Y}(T_{2},\widehat{\chi}_{Y}(T_{3}\right)$. We only considered up to four swaps. We detect zero Young diagram pairs with this degeneracy having less than two swaps.}
  \label{Tab:T2T3data}
\end{table}

\begin{table}
  \centering 
  \begin{tabular}{|c|c|c|c|c|c|c|}
\hline
 $n$  & 0 swaps & 1 swap  & 2 swaps & 3 swaps & 4 swaps & 5 swaps  \\
 \hline
  24 & 0 & 0 & 0 & 1 &0 & 0 \\
  \hline
  25 & 0 & 0 & 0 & 0 & 0 &0\\
  \hline
  26 & 0 & 0 & 0 & 0 &0 &0\\
  \hline
  27 & 0 & 0 & 0 & 2 &0 &0\\
  \hline
  28 & 0 & 0 & 0 & 2 &0 &0\\
  \hline
  29 & 0 & 0 & 0 & 0 &0 &0\\
  \hline
  30 & 0 & 0 & 0 & 4 & 0 & 0\\
  \hline
  31 & 0 & 0 & 0 & 3 & 0 &0 \\
  \hline
  32 & 0 & 0 & 0 & 14 & 0 &0 \\
  \hline
  33 & 0 & 0 & 0 & 15 & 0 &0 \\
  \hline
  34 & 0 & 0 & 0 & 16 & 0 &0 \\
  \hline
  35 & 0 & 0 & 0 & 33 & 0 &0 \\
  \hline
  36 & 0 & 0 & 0 & 44 & 0 &0 \\
  \hline
  37 & 0 & 0 & 0 & 44 & 0 & 0\\
  \hline
  38 & 0 & 0 & 0 & 62 & 0 &0 \\
  \hline
  39 & 0 & 0 & 0 & 75 & 0 &0 \\
  \hline
  40 & 0 & 0 & 0 & 123 & 0 &0 \\
  \hline
  41 & 0 & 0 & 0 & 135 & 0 & 0 \\
  \hline
  42 & 0 & 0 & 0 & 186 & 2 & 0 \\
  \hline
  43 & 0 & 0 & 0 & 218 & 2 & 0 \\
  \hline
  44 & 0 & 0 & 0 & 298 & 6 &  2 \\
  \hline
  45 & 0 & 0 & 0 & 394 & 8 &  0 \\
  \hline
  46 & 0 & 0 & 0 & 484 & 8 &  4 \\
  \hline
  47 & 0 & 0 & 0 & 588 & 6 &  2 \\
  \hline
  48 & 0 & 0 & 0 & 797 & 20 &  5 \\
  \hline
\end{tabular}
  \caption{This table measures $k_{swap}$ for all pairs of diagrams having degenerate $\left(\widehat{\chi}_{Y}(T_{2},\widehat{\chi}_{Y}(T_{3}, \widehat{\chi}_{Y}(T_{4}\right)$. We only considered up to five swaps. We detect zero Young diagram pairs with this degeneracy having less than three swaps.}
  \label{Tab:T2T3T4data}
\end{table}
\begin{table}
  \centering 
  \begin{tabular}{|c|c|c|c|c|c|c|}
\hline
 $n$  & 0 swaps & 1 swap  & 2 swaps & 3 swaps & 4 swaps & 5 swaps  \\
 \hline
  42 & 0 & 0 & 0 & 0 & 2 & 0 \\
  \hline
  43 & 0 & 0 & 0 & 0 & 2 & 0 \\
  \hline
  44 & 0 & 0 & 0 & 0 & 6 &  0 \\
  \hline
  45 & 0 & 0 & 0 & 0 & 8 &  0 \\
  \hline
  46 & 0 & 0 & 0 & 0 & 8 &  0 \\
  \hline
  47 & 0 & 0 & 0 & 0 & 6 &  0 \\
  \hline
  48 & 0 & 0 & 0 & 0 & 20 &  0 \\
  \hline
  49 & 0 & 0 & 0 & 0 & 24 &  0 \\
  \hline
  50 & 0 & 0 & 0 & 0 & 28 &  0 \\
  \hline
  51 & 0 & 0 & 0 & 0 & 30 &  0 \\
  \hline
  52 & 0 & 0 & 0 & 0 & 40 &  0 \\
  \hline
  53 & 0 & 0 & 0 & 0 & 50 &  0 \\
  \hline
  54 & 0 & 0 & 0 & 0 & 74 &  0 \\
  \hline
  55 & 0 & 0 & 0 & 0 & 90 &  0 \\
  \hline
  56 & 0 & 0 & 0 & 0 & 116 &  0 \\
  \hline
  57 & 0 & 0 & 0 & 0 & 140 &  0 \\
  \hline
  58 & 0 & 0 & 0 & 0 & 182 &  0 \\
  \hline
  59 & 0 & 0 & 0 & 0 & 230 &  0 \\
  \hline
  60 & 0 & 0 & 0 & 0 & 309 &  0 \\
  \hline
  61 & 0 & 0 & 0 & 0 & 340 &  0 \\
  \hline
  62 & 0 & 0 & 0 & 0 & 420 &  0 \\
  \hline
  63 & 0 & 0 & 0 & 0 & 534 &  0 \\
  \hline
  64& 0 & 0 & 0 & 0 & 666 &  0 \\
  \hline
  65& 0 & 0 & 0 & 0 & 800 &  0 \\
  \hline
\end{tabular}
  \caption{This table measures $k_{swap}$ for all pairs of diagrams having degenerate $\left(\widehat{\chi}_{Y}(T_{2},\widehat{\chi}_{Y}(T_{3}, \widehat{\chi}_{Y}(T_{4} , \widehat{\chi}_{Y}(T_{5}  \right)$. We considered up to five swaps. We detect zero Young diagram pairs with this degeneracy having less than four swaps.}
  \label{Tab:T2T3T4T5data}
\end{table}
In this last paragraph, we discuss the case of $k_{deg} = 6$. The first instance of partitions appearing for which $\{\widehat{\chi}_{Y}(T_{2}) ,  \widehat{\chi}_{Y}(T_{3}) ,  \widehat{\chi}_{Y}(T_{4}) , \widehat{\chi}_{Y}(T_{5}) , \widehat{\chi}_{Y}(T_{6}) \}$ are degenerate is at $n = 80$. There were 3 pairs of diagrams having degenerate values for the above list. They are 
\begin{eqnarray}
\label{eq:3a}
	Y_{1} &=& (16, 12, 12, 10, 4, 4, 4, 4, 4, 3, 3, 1, 1, 1, 1),\\
\label{eq:3b}
	 Y'_{1} &=& (15, 15, 10, 7, 7, 7, 3, 3, 3, 2, 2, 2, 2, 2),\\
	  \nonumber\\
\label{eq:3c}
	Y_{2} &=& (16, 12, 12, 8, 4, 4, 4, 4, 3, 3, 3, 3, 1, 1, 1, 1) ,\\
\label{eq:3d}
	Y'_{2} &=& (15, 15, 9, 7, 6, 6, 4, 3, 3, 2, 2, 2, 2, 2, 2) \\
	\nonumber\\
\label{eq:3e}
	Y_{3} &=& (15, 11, 11, 9, 4, 4, 4, 4, 4, 4, 3, 3, 1, 1, 1, 1) ,\\
\label{eq:3f}
	Y'_{3} &=& (14, 14, 9, 6, 6, 6, 6, 3, 3, 3, 2, 2, 2, 2, 2).
\end{eqnarray}
There are five swaps for each pair. For instance, the symbol of $(Y_{2}, Y'_{2})$ is
\bea
\label{eq:3g}
(<,>,<,<,=,>,>,>,>,=,<,<,<,=,>,=).
\eea
The data presented in tables \ref{Tab:T2data}, \ref{Tab:T2T3data}, \ref{Tab:T2T3T4data}, \ref{Tab:T2T3T4T5data}, and in equation (\ref{eq:3g}) indicates that for a pair of Young diagrams of $n$ boxes having a given $k_{deg}$, there is a minimum number of swaps of the inequality sign in the row sums for the pairs. This data is summarized in Table \ref{Tab:Datasummary} and is in support of our conjecture. For all partition pairs having $k_{deg} = 2$ in our data, there are zero pairs having less than one swap. For all partition pairs having $k_{deg} = 3$ in our data, there are zero pairs having less than two swaps and so on.  
\begin{table}
  \centering 
  \begin{tabular}{|c|c|}
\hline
 $k_{deg}$ & Minimum number of swaps\\
 \hline
  2 & 1 \\
\hline
  3 & 2 \\
  \hline
  4 & 3 \\
  \hline
  5 & 4 \\
  \hline
  6 & 5 \\
  \hline
\end{tabular}
  \caption{Table showing the minimum number of swaps in the direction of the inequality sign in the row sum relations of partitions with $k_{deg} = 2,3,4,5,6$.}
  \label{Tab:Datasummary}
\end{table}

\section{From content polynomials to power sums}\label{contpolypowersums}

\subsection{Content polynomials}

An important result in \cite{CGS} relates the normalized characters of the irreps $Y$ of $S_{n}$ to content polynomials $\langle c^{p}(Y) \rangle$. The content polynomials are essentially power sums evaluated on the contents of the Young diagram $Y$. The contents of an irrep $Y$, represented as a Young diagram of $n$ boxes are the numbers $j-i$, where $(i,j)$ are the positions of the boxes in the diagram. For example, for the diagram $(4,2,2,1)$, the contents are
\bea
\nonumber
\ytableausetup{boxsize=1.8em}
\begin{ytableau}
       \none &0&1&2&3 \\
  \none  &-1&0 \\
  \none &-2&-1\\
  \none &-3
\end{ytableau}.
\eea
Thus, the content polynomial $\langle c^{p}(Y) \rangle$ for $Y = (4,2,2,1)$ is simply
\bea
\label{eq:5a0}
	\langle c^{p}(Y) \rangle = (0)^{p} + (1)^{p} + (2)^{p} + (3)^{p} + (-1)^{p} + (0)^{p} + (-2)^{p} + (-1)^{p} + (-3)^{p}. 
\eea
The normalized characters of the cycle operators, $T_{k}$, in irrep $Y$ may be expressed in terms of content polynomials in the following way \cite{CGS}
\begin{equation}
	\widehat{\chi}_{Y}\left(T_{k}\right) = \langle c^{k-1}(Y) \rangle + \sum\limits^{k-2}_{l=0} \sum\limits_{\lambda \vdash l} \Omega^{\lambda}_{k-1}(n) \langle c^{\lambda_{1}}(Y) \rangle \langle c^{\lambda_{2}}(Y) \rangle \cdots \langle c^{\lambda_{r}}(Y) \rangle,
	\label{eq:5a}
\end{equation}
where the $\Omega^{\lambda}_{k-1}(n)$ is a polynomial in $n$ and is independent of $Y$, and $\lambda$ is a partition (and an irrep) of $l$, written as $\lambda = (\lambda_{1} , \lambda_{2} , \cdots , \lambda_{r} )$. From (\ref{eq:5a}), we see that, for $k=2$, $\widehat{\chi}_{Y}\left(T_{2}\right) = \widehat{\chi}_{Y'}\left(T_{2}\right)$ if and only if $ \langle c(Y) \rangle =  \langle c(Y') \rangle$. Furthermore, $\{ \widehat{\chi}_{Y}\left(T_{2}\right) , \widehat{\chi}_{Y}\left(T_{3}\right) \} = \{\widehat{\chi}_{Y'}\left(T_{2}\right) , \widehat{\chi}_{Y'}\left(T_{3}\right) \}$ if and only if $\{ \langle c(Y) \rangle, \langle c^{2}(Y) \rangle \}=  \{ \langle c(Y') \rangle  , \langle c^{2}(Y') \rangle\}$. To see this, for $k=3$,
\bea
	 \widehat{\chi}_{Y}\left(T_{3}\right) &=& \langle c^{2}(Y) \rangle + \sum\limits^{1}_{l=0} \sum\limits_{\lambda \vdash l} \Omega^{\lambda}_{2}(n) \langle c^{\lambda_{1}}(Y) \rangle \nonumber \\
\label{eq:5b}
	 	&=& \langle c^{2}(Y) \rangle + \Omega^{1}_{2}(n)  \langle c(Y) \rangle.
\eea
Thus, for the pair $(Y,Y')$, if the normalized characters of $T_{2}$ are degenerate, which implies the first content polynomials $\langle c(Y) \rangle$ are degenerate, then the normalized characters of $T_{3}$ are degenerate if and only if the second content polynomial $ \langle c^{2}(Y) \rangle$ is degenerate, recalling that the polynomial $\Omega$ only depends on $k,\lambda$ and $n$, and not on $Y$. For $k=4$,
\bea
	 \widehat{\chi}_{Y}\left(T_{4}\right) &=& \langle c^{3}(Y) \rangle + \sum\limits^{2}_{l=0} \sum\limits_{\lambda \vdash l} \Omega^{\lambda}_{3}(n) \langle c^{\lambda_{1}}(Y) \rangle \langle c^{\lambda_{2}}(Y) \rangle \nonumber \\
\label{eq:5c}
	 	&=& \langle c^{3}(Y) \rangle + \Omega^{1}_{3}(n)  \langle c(Y) \rangle  + \Omega^{(2)}_{3}(n)  \langle c^{2}(Y) \rangle + \Omega^{(1,1)}_{3}(n)  \langle c(Y) \rangle \langle c(Y) \rangle .
\eea
Again we see that degeneracy in $ \widehat{\chi}_{Y}\left(T_{2}\right) $ and $ \widehat{\chi}_{Y}\left(T_{3}\right)$, which implies degeneracy in $ \langle c(Y) \rangle $ and $ \langle c^{2}(Y) \rangle$, implies that $ \widehat{\chi}_{Y}\left(T_{4}\right)$ is degenerate if and only if $ \langle c^{3}(Y) \rangle$ is degenerate for $Y$ and $Y'$. For higher $k$, a proof may be constructed by induction on $k$. For the partition pair, $(Y,Y')$, if their normalized characters for $\{ T_{2} , T_{3} , \cdots , T_{k} \}$ are degenerate, which implies that their content polynomials $\{ \langle c \rangle , \langle c^{2} \rangle , \cdots , \langle c^{k-1} \rangle \}$ are degenerate, then their $T_{k+1}$ normalized character is degenerate if and only if their content polynomial $\langle c^{k} \rangle$ is degenerate. Thus, we may shift our focus from studying the normalized characters of the cycle operators in irrep $Y$ to instead studying the content polynomials evaluated on $Y$.

\subsection{Power sums}

The power sum polynomial of degree $q$ evaluated on variables $x_{1}, x_{2}, \cdots, x_{l}$, is defined by \cite{McD}
\bea
\label{eq:5ca}
p_{q}(x_{1}, \cdots , x_{l}) = \sum\limits^{l}_{i=1} x^{q}_{i}.
\eea
Letting $x_{1} = 1, x_{2} = 2,\cdots, x_{l} = l$, $p_{q}$ defines an integer polynomial $g$ of degree $q+1$ depending on $l$:
\bea
\label{eq:5cb}
	p_{q}(1, \cdots, l) = \sum\limits^{l}_{i=1} i^{q} = g_{q+1}(l).
\eea
Furthermore, $g_{q+1}(l)$ may be expanded in terms of powers $l^{\alpha}, \alpha = 1, 2, \cdots, q+1$ with some expansion coefficients $a_{\alpha}$,
\bea
\label{eq:5cc}
	g_{q+1}(l) = \sum\limits^{q+1}_{\alpha} a_{\alpha}\,(l)^{\alpha}.
\eea
For the first few $q$,
\bea
	\sum\limits^{l}_{i=1} i^{1} &=& g_{2}(l) = \frac{l}{2} + \frac{l^{2}}{2} \nonumber\\
	\sum\limits^{l}_{i=1} i^{2} &=& g_{3}(l) = \frac{l}{6} + \frac{l^{2}}{2} +  \frac{l^{3}}{3}
\label{eq:5e}\\
	\sum\limits^{l}_{i=1} i^{3} &=& g_{4}(l) = \frac{l^{2}}{4} + \frac{l^{3}}{2} + \frac{l^{4}}{4} \nonumber\\
	\sum\limits^{l}_{i=1} i^{4} &=& g_{5}(l) = -\frac{l}{30} + \frac{l^{3}}{3} + \frac{l^{4}}{2} + \frac{l^{5}}{5}.\nonumber
\eea
Now consider the content polynomials again. Denote by $ \langle c^{q}(y_{j}) \rangle $ the sum of the contents $c$ in row $j$ of $Y$ raised to the power $q$. The contributions to $\langle c^{q}(y_{j}) \rangle$ from contents $1,2, \cdots , y_{j}-j$ is, from (\ref{eq:5cb}) and (\ref{eq:5cc}), 
\bea
\label{eq:5d}
	\sum\limits^{y_{j}-j}_{i=1} i^{q} = g_{q+1}(y_{j}-j) =  \sum\limits^{q+1}_{\alpha=1} a_{\alpha} (y_{j}-j)^{\alpha}.
\eea 
Similarly, the contributions to $\langle c^{q}(y_{j}) \rangle$ from $-j+1 , -j+2 , \cdots, -1$ is a polynomial in $j$ also of degree $q+1$. Denote this polynomial by $p^{-}_{q+1}(j)$. For $\langle c^{q}(y_{j}) \rangle$, we add these two polynomials together. We first note that for $q=0$, $\langle c^{0}(y_{j})\rangle = y_{j}$, the number of boxes in row $j$, as expected. The power sums of the contents of row $j$ in diagram $Y$ are
\bea
\label{eq:5g}
	 \langle c^{q}(y_{j}) \rangle = p^{-}_{q+1}(j) + \sum\limits^{q+1}_{\alpha=1} a_{\alpha} (y_{j}-j)^{\alpha}.
\eea
To obtain the content polynomial over full diagram $Y$, we sum the contribution from each row. Let $l(Y)$ denote the number of rows present in $Y$. Then
\bea
\label{eq:5h}
	\langle c^{q}(Y)\rangle = \sum\limits^{l(Y)}_{j=1} \langle c^{q}(y_{j}) \rangle = \sum\limits^{l(Y)}_{j=1} \left[ p^{-}_{q+1}(j) + \sum\limits^{q+1}_{\alpha=1} a_{\alpha} (y_{j}-j)^{\alpha} \right].
\eea
We want to compare two content polynomials, for a given $q$, from two distinct partitions $Y$ and $Y'$. 
If the length of $Y$ is longer than that of $Y'$, we pad out the remaining entries in $Y'$ with 0's. For example, $Y= (4, 3, 2, 2, 1)$ and $Y' = (5, 4, 3)$, are two partitions of 12, one having 5 rows and the other 3 rows. We then write $Y' = (5, 4, 3,0,0)$. Appending these extra 0's in a partition will not change its content polynomial. Thus, when comparing $\langle c^{q}(Y)\rangle$ and $\langle c^{q}(Y')\rangle$, each is given by formula (\ref{eq:5h}) and $l(Y) = l(Y')$. 
This leads us to only compare the polynomial $p_{q+1}(y_{j}-j) $ in (\ref{eq:5h}). The coefficients $a_{\alpha}$ do not depend on values of the row lengths of $Y$. Thus, the $a_{\alpha}$ are equivalent for any two $Y$ and $Y'$.
Since $ \langle c^{0}(y_{j}) \rangle$ simply counts the number of boxes in row $j$ for diagram $Y$, $\langle c^{0}(Y)\rangle$ is the total number of boxes in $Y$. Both $Y$ and $Y'$ have $n$ boxes, which means
\bea
\label{eq:5i}
	 \sum\limits^{l(Y)}_{j=1} (y_{j} - j) =  \sum\limits^{l(Y')}_{j=1} (y'_{j} - j).
\eea
For $q=1$, the content polynomials $\langle c(Y)\rangle $ and $\langle c(Y')\rangle $ are equal if and only if
\bea
\label{eq:5j}
	 \sum\limits^{l(Y)}_{j=1} (y_{j} - j)^{2} =  \sum\limits^{l(Y')}_{j=1} (y'_{j} - j)^{2}.
\eea
For $q=2$, and assuming $\langle c(Y)\rangle = \langle c(Y')\rangle $, the next content polynomials are equal, $\langle c^{2}(Y)\rangle = \langle c^{2}(Y')\rangle$, if and only if 
\bea
\label{eq:5k}
	 \sum\limits^{l(Y)}_{j=1} (y_{j} - j)^{3} =  \sum\limits^{l(Y')}_{j=1} (y'_{j} - j)^{3}.
\eea
For general $q$, assuming $\left\{ \langle c(Y)\rangle ,\langle c^{2}(Y)\rangle , \cdots ,\langle c^{q-1}(Y)\rangle  \right\} = \left\{ \langle c(Y')\rangle ,\langle c^{2}(Y')\rangle , \cdots ,\langle c^{q-1}(Y')\rangle \right\} $, then we have $\langle c^{q}(Y)\rangle = \langle c^{q}(Y')\rangle$ if and only if
\bea
\label{eq:5l}
	 \sum\limits^{l(Y)}_{j=1} (y_{j} - j)^{q+1} =  \sum\limits^{l(Y')}_{j=1} (y'_{j} - j)^{q+1}.
\eea
In the rest of this work, we denote the difference $ \sum\limits_{j} \left[ (y_{j} - j)^{q} - (y'_{j} - j)^{q} \right] $ by $\Delta^{(Y,Y')}_{q}$. Note immediately that $\Delta^{(Y,Y')}_{q} = 0$ for $q = 0,1$, and that the order of $Y,Y'$ in $\Delta^{(Y,Y')}_{q}$ matters: $\Delta^{(Y,Y')}_{q} = -\Delta^{(Y',Y)}_{q}$.

In the previous section, we have argued that the list of normalized characters $\{ \widehat{\chi}_{Y}(T_{2}) , \cdots , \widehat{\chi}_{Y}(T_{k})   \}$ for $Y,Y'$ are degenerate if and only if the list of content polynomials $\{ \langle c(Y) \rangle , \cdots ,  \langle c^{k-1}(Y) \rangle \}$ are degenerate for $Y,Y'$. What we have shown in this section is that this occurs if and only if 
\bea
\label{eq:5m}
\Delta^{(Y,Y')}_{q} = 0, \hspace{20pt} q = 1 , 2 , \cdots , k.
\eea
As an example, consider again the pair $(Y_{2} , Y'_{2})$ which were Young diagrams of 80 boxes. In our data, we found $\{ T_{2} , T_{3} , T_{4} , T_{5} , T_{6} \}$ were degenerate for this pair. Checking the content polynomials and power sums for this pair, we find, as expected
\bea
	\langle c(Y_{2}) \rangle &=& \langle c(Y'_{2}) \rangle = 0, \hspace{20pt} \langle c^{2}(Y_{2}) \rangle = \langle c^{2}(Y'_{2}) \rangle = 3880, \hspace{20pt} \langle c^{3}(Y_{2}) \rangle = \langle c^{3}(Y'_{2}) \rangle = 0,\nonumber \\
	  \langle c^{4}(Y_{2}) \rangle &=& \langle c^{4}(Y'_{2}) \rangle = 438664, \hspace{20pt} \langle c^{5}(Y_{2}) \rangle =  \langle c^{5}(Y'_{2}) \rangle, \hspace{20pt} \langle c^{6}(Y_{2}) \rangle >  \langle c^{6}(Y'_{2}) \rangle, \nonumber\\
	  &&\nonumber \\
	\Delta^{(Y_{2} , Y'_{2})}_{2} &=& 0, \hspace{20pt} \Delta^{(Y_{2} , Y'_{2})}_{3} = 0, \hspace{20pt} \Delta^{(Y_{2} , Y'_{2})}_{4} = 0,\nonumber \\  
	\Delta^{(Y_{2} , Y'_{2})}_{5} &=& 0, \hspace{20pt} \Delta^{(Y_{2} , Y'_{2})}_{6} = 0, \hspace{20pt} \Delta^{(Y_{2} , Y'_{2})}_{7} > 0.\nonumber
\eea

\section{Analytic arguments supporting the conjecture}\label{analarguments}

The goal in this section is to offer some analytic evidence in support of our conjecture. Studying the functions $\Delta^{(Y,Y')}_{k}$, we first write them in terms of the completely symmetric homogeneous polynomials $h_{k-1}$. Then we use Karamata's theorem \cite{Karamata} to show that if $k^{(Y,Y')}_{swap} = 0$, then $\Delta^{(Y,Y')}_{2} > 0$. This result implies that the normalized characters of the first cycle operator $T_{2}$ for $Y$ and $Y'$ are distinct. The conclusion we draw from this is that in order for $\Delta^{(Y,Y')}_{2} =0$, we require $k^{(Y,Y')}_{swap} \geq 1$. Then, we assume $k^{(Y,Y')}_{swap} = 1$, and $\Delta^{(Y,Y')}_{2} = 0$, and proceed to show that $\Delta^{(Y,Y')}_{3} > 0$. The conclusion we draw from this is that in order for both $\Delta^{(Y,Y')}_{2} = \Delta^{(Y,Y')}_{3} = 0$, we require $k^{(Y,Y')}_{swap} \geq 2$. Lastly, restricting ourselves to a simple case in which $Y'$ is obtained from $Y$ by moving three boxes, assuming $k^{(Y,Y')}_{swap} = 2$, and assuming $\Delta^{(Y,Y')}_{2} = \Delta^{(Y,Y')}_{3} = 0$, we show that $\Delta^{(Y,Y')}_{4}>0$. The three-box move that we explain in section \ref{kswapeq2} is the simplest way to produce $k^{(Y,Y')}_{swap} = 2$.

While we are not able to prove our conjecture in full generality, we hope that the expressions and results we give in this section will be useful in studying this problem. 

\subsection{Expressing $\Delta_{k}$ in terms of the completely symmetric homogeneous polynomials}

Let $\widetilde{Y}$ denote the Young diagram $Y$ with the shifted row lengths
\bea
	\wt{Y} = (\wt{y}_{1} , \cdots , \wt{y}_{d}), \hspace{20pt} \wt{y}_{j} = y_{j} - j.
\eea
In general, we can express $\Delta^{(Y,X)}_{k}$ in the following way,
\begin{eqnarray}
\label{eq:Sec5.1a}
	\Delta^{(Y,X)}_{k} &=& \sum\limits^{d}_{i=1} \left( \wt{y}^{k}_{i} - \wt{x}^{k}_{i} \right) = \sum\limits^{d}_{i=1} h_{k-1}\left( \wt{y}_{i} , \wt{x}_{i} \right) \left(\wt{y}_{i} - \wt{x}_{i}  \right),
\end{eqnarray}
where the $h_{k-1}\left( \wt{y}_{i} , \wt{x}_{i} \right)$ are the completely symmetric homogeneous polynomials in two variables, $ \wt{y}_{i} , \wt{x}_{i}$. For the first few $k$, we have
\begin{eqnarray}
	h_{1}\left( \wt{y}_{i} , \wt{x}_{i} \right) &=& \wt{y}_{i} + \wt{x}_{i}. \nonumber \\
	\label{eq:Sec5.1b}
	h_{2}\left( \wt{y}_{i} , \wt{x}_{i} \right) &=& \wt{y}^{2}_{i} + \wt{y}_{i} \wt{x}_{i} + \wt{x}^{2}_{i}. \\
	h_{3}\left( \wt{y}_{i} , \wt{x}_{i} \right) &=& \wt{y}^{3}_{i} + \wt{y}^{2}_{i}\wt{x}_{i} + \wt{y}_{i} \wt{x}^{2}_{i} + \wt{x}^{3}_{i}. \nonumber \\
	h_{4}\left( \wt{y}_{i} , \wt{x}_{i} \right) &=& \wt{y}^{4}_{i} + \wt{y}^{3}_{i}\wt{x}_{i} + \wt{y}^{2}_{i} \wt{x}^{2}_{i} + \wt{y}_{i} \wt{x}^{3}_{i} + \wt{x}^{4}_{i} . \nonumber
\end{eqnarray}
Since $\wt{y}_{j}$ and $\wt{x}_{j}$ are the shifted row lengths, we have $\left(\wt{y}_{i} - \wt{x}_{i}  \right) = \left(y_{i}-x_{i}  \right)$.

Furthermore, it is also known that $h_{k}$ can be written in terms of $h_{1} , h_{2} , \cdots , h_{k-1}$ and power sum polynomials \cite{McD},
\bea
\label{eq:Sec5.1c}
	h_{k} (\wt{y}_{i} , \wt{x}_{i}) = \frac{1}{k} \sum\limits^{k-1}_{\alpha=0} h_{\alpha}( \wt{y}_{i} , \wt{x}_{i})p_{k-\alpha}( \wt{y}_{i} , \wt{x}_{i}).
\eea
Note also that some parts of $\wt{X}$ and $\wt{Y}$ may be negative. However, without affecting our analysis, we may shift both $\wt{X}$ and $\wt{Y}$ by a constant $\alpha$ so that both have only positive (or possibly zero) parts. To see this, shift both $\wt{X}$ and $\wt{Y}$ by a constant $\alpha$, to form $\wt{X}^{(\alpha)} = \wt{X}+\alpha , \wt{Y}^{(\alpha)} = \wt{Y} + \alpha$. Then the new $\Delta_{k}$ becomes
\bea
	\Delta^{(Y^{(\alpha)},X^{(\alpha)})}_{k} &=& \sum\limits^{d}_{i=1}\left[ \left( \widetilde{y}_{i}+\alpha\right)^{k}  - \left( \widetilde{x}_{i}+\alpha\right)^{k}  \right]\\
	&=&  \sum\limits^{d}_{i=1} \left[ \sum\limits^{k}_{l=0} \binom{k}{l}\alpha^{k-l} \widetilde{y}_{i}^{l} - \sum\limits^{k}_{l=0} \binom{k}{l}\alpha^{k-l} \widetilde{x}_{i}^{l} \right] \\
	&=& \sum\limits^{k}_{l=0}  \binom{k}{l} \alpha^{k-l} \;  \sum\limits^{d}_{i=1} \left[ \widetilde{y}_{i}^{l} - \widetilde{x}_{i}^{l} \right]\\
	&=& \sum\limits^{k}_{l=0}  \binom{k}{l} \alpha^{k-l} \Delta^{(Y,X)}_{l}.
\eea
Recall that if $X$ and $Y$ are Young diagrams both of $n$ boxes, then $\Delta^{(Y,X)}_{1} = 0$. For $k=2$, since, $\Delta^{(Y,X)}_{1} = 0$, we see that $\Delta^{(Y^{(\alpha)},X^{(\alpha)})}_{2} = \Delta^{(Y,X)}_{2}$. For $k=3$, if $\Delta^{(Y,X)}_{1} = \Delta^{(Y,X)}_{2} = 0$, then $\Delta^{(Y^{(\alpha)},X^{(\alpha)})}_{3} = \Delta^{(Y,X)}_{3}$. In general, if $\Delta^{(Y,X)}_{1} = \Delta^{(Y,X)}_{2} = \cdots = \Delta^{(Y,X)}_{l-1}  = 0  $, then $\Delta^{(Y,X)}_{l} $ is invariant under the shift:
\bea
	\Delta^{(Y^{(\alpha)},X^{(\alpha)})}_{l}  = \Delta^{(Y,X)}_{l} .
\eea
Thus, we can assume that $\wt{Y}$ and $\wt{X}$ have non-negative parts: $\wt{y}_{1} > \wt{y}_{2} > \cdots > \wt{y}_{d} \geq 0$, and similarly for $\wt{X}$ without affecting our results.

\subsection{$k^{(Y,X)}_{swap} = 0$ case}

We assume $X \prec Y$, where $\wt{Y}$ is such that $\wt{y}_{1} > \wt{y}_{2} > \cdots > \wt{y}_{d} \geq 0$, and similarly for $\wt{X}$. We also assume that $X\neq Y$. The inequality sign in the row sum relations only faces in the direction of the $Y$ row sums and does not swap direction. It is indeed known \cite{Diaconis} that for this case, the normalized character of $T_{2}$ in irrep $Y$ is greater than that in irrep $X$. In our formulation, this should mean that $\Delta^{(Y,X)}_{2} > 0$. We apply Karamata's theorem, whose steps are explained below, to the case of completely symmetric homogeneous polynomials to show here that indeed $\Delta^{(Y,X)}_{2} > 0$ for all $X\prec Y$.

Karamata's argument begins by defining $A_{i} = y_{1} + \cdots + y_{i}$, and $B_{i} = x_{1} + \cdots + x_{i}$. Then,
\bea
	\left(y_{i}-x_{i}  \right) = (A_{i} - A_{i-1}) - (B_{i} - B_{i-1}) = (A_{i} - B_{i}) - (A_{i-1} - B_{i-1}) \nonumber.
\eea
The expression for $\Delta^{(Y,X)}_{k}$ in (\ref{eq:Sec5.1a}) now becomes 
\bea
\label{eq:sec5.2a}
	\Delta^{(Y,X)}_{k} = \sum\limits^{d-1}_{i=1} \big[h_{k-1}\left( \wt{y}_{i} , \wt{x}_{i} \right) - h_{k-1}\left( \wt{y}_{i+1} , \wt{x}_{i+1} \right) \big] \left( A_{i} - B_{i} \right).
\eea
In general, the completely symmetric homogeneous polynomials $h_{k}(x_{1},x_{2},\cdots , x_{d})$ for $x_{i} > x_{i+1} $ and $ x_{d} \geq 0 $ are Schur convex functions. This means that if $X\prec Y$, then $h_{k}(x_{1} , \cdots , x_{d}) < h_{k}(y_{1} , \cdots , y_{d})$\footnote{See also \cite{TerryTao} for an informal but informative discussion on completely symmetric homogeneous polynomials.}. In equation (\ref{eq:sec5.2a}) since $\wt{y}_{i} > \wt{y}_{i+1} \geq 0$ and $\wt{x}_{i} > \wt{x}_{i+1} \geq 0$ for all $i = 1, 2, \cdots , d-1$, the polynomial 
\bea
\label{eq:sec5.2b}
h_{k-1}\left( \wt{y}_{i} , \wt{x}_{i} \right) > h_{k-1}\left( \wt{y}_{i+1} , \wt{x}_{i+1} \right).
\eea
From (\ref{eq:sec5.2b}) we learn that the coefficients of $(A_{i} - B_{i})$ in (\ref{eq:sec5.2a}) are always positive. Furthermore, since $X\prec Y$, we have $A_{i} \geq B_{i}$, where we have strict inequality for some $i$. Thus, following the reasoning in the Karamata theorem for the $h_{k}$ polynomials, we conclude that $\Delta^{(Y,X)}_{k} > 0$, which includes the case $\Delta^{(Y,X)}_{2} > 0$.

As a corollary to this, if $\Delta^{(Y,X)}_{2}$ is to be equal to zero, we require, for some $i$, $A_{i} < B_{i}$. This means that $X$ and $Y$ are incomparable according to the majorization ordering and the row sums described in (\ref{eq:1a}). In other words, the inequality sign in the row sums of $X$ and $Y$ needs to switch direction at least once.

\subsection{$k^{(Y,X)}_{swap} = 1$ case}

In this section we argue that for the Young diagram pair $(Y_{F},Y_{I})$ that we define below, if $k^{(Y_{F},Y_{I})}_{swap} = 1$ and $\Delta^{(Y_{F},Y_{I})}_{2} = 0$, we must have $\Delta^{(Y_{F},Y_{I})}_{3} > 0$.

We first define a one-box move. Consider a Young diagram $Y$ of $n$ boxes. The lengths of the rows of $Y$ are $(y_{1}, y_{2}, \cdots , y_{d} )$ where we require $y_{i}\geq y_{i+1}$ for $i=1,2,\cdots,d-1$, and $y_{d}\geq 0$. A one-box move is defined to be the movement of a single box from the corner of some row $i$ in $Y$ to the corner of some other row $j$. If $i>j$ then this corresponds to an upward movement of the box. Conversely, if $i<j$ the movement of the box is downward. A one-box move is only possible if the result $Y' = (y'_{1}, \cdots , y'_{d})$ is still a valid Young diagram. That is, $Y'$ must obey the same constraints as those for $Y$, as described above. For example, if $i>j$, then we must have
\bea
	Y' &=& (y'_{1} , \cdots , y'_{j} , \cdots , y'_{i} , \cdots , y'_{d}), \\
		&=& (y_{1} , \cdots , y_{j}+1 , \cdots , y_{i}-1 , \cdots , y_{d}),\\
		&\mathrm{where}& y_{1} \geq  \cdots \geq y_{j-1} \geq y_{j}+1 \geq y_{j+1} \geq \cdots y_{i-1} \geq y_{i}-1 \geq y_{i+1} \geq \cdots .
\eea
As an example, consider the Young diagram $(4,2,2,1)$:
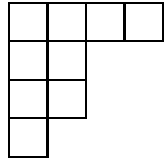
\bea
\nonumber
\ytableausetup{boxsize=1.2em}
\begin{ytableau}
       \none &&&& \\
  \none  &&\\
  \none &&\\
  \none &
\end{ytableau}
\eea
The movement of the box at the corner of row 1 down to the corner of row 2 is permitted, while its movement to the corner of row 3 is not. This box is also permitted to move to the corner of row 4 and also to the corner of row 5 with that box being the only box in row 5. The corner box in row 2 cannot move either up or down. The corner box in row 3 is permitted to move up to row 1, or to row 2, while it is not permitted to move down to row 4. It is possible for this box to move down to row 5. Lastly, the corner box in row 4 can move up to rows 1 and 2 only. 

We consider some initial Young diagram $Y_{I}$ with $n$ boxes and $l(Y_{I})$ rows. Consider row $s$ in $Y_{I}$, where $1 < s < l(Y_{I})$. This divides $Y_{I}$ into two regions - the region above row $s$ and the region below and including row $s$. One can form another Young diagram $Y_{F}$ from $Y_{I}$ with a series of one-box moves. One may move boxes from row $s$, and from rows underneath $s$, downward to rows further below. After this series of downward moves, denote this intermediate Young diagram by $Y_{int}$. Thereafter, one may move boxes from row $s$, and from rows above $s$, upward to rows further above. In this way $Y_{I} \prec Y_{F}$ for the rows above row $s$, and then $Y_{I} \succ Y_{F}$ for row $s$ and below\footnote{Technically we mean weak majorization here. This means that we do not require equality when the final row in the top region (above row $s$) is summed in the row sum relations. In the examples we present below, $s=2$ for $Y=(3,3)$ and $Y'=(4,1,1)$. The row sums for $(Y,Y')$ above row $s$ give 3 for $Y$ and 4 for $Y'$. However, for $Y = (4,2,2,1)$ and $Y' = (5,1,1,1,1)$, $s = 3$ and the row sums above row $s$ both give 6 for $Y$ and $Y'$. Our results incorporate both these cases.}. This is schematically shown in figure \ref{fig1aa}.
\begin{center}
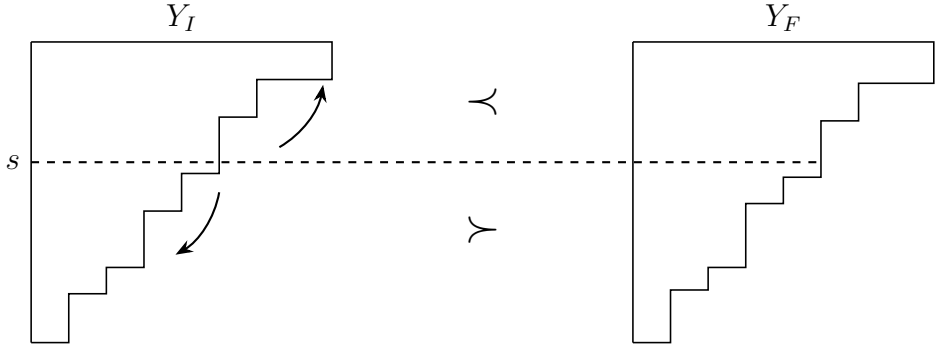
\begin{figure}[h!]
\begin{tikzpicture}[semithick, >=Stealth]
\coordinate (A0) at (-6+3,0);
\coordinate (A1) at (-2+3,0);
\coordinate (Abottomleft) at (-6+3,-4);
\draw (A0) -- (Abottomleft);
\coordinate (A2) at (-2+3,-0.5);
\coordinate (A3) at (-3+3,-0.5);
\coordinate (A4) at (-3+3,-1);
\coordinate (A5) at (-3.5+3,-1);
\coordinate (A6) at (-3.5+3,-1.75);
\coordinate (A7) at (-4+3,-1.75);
\coordinate (A8) at (-4+3,-2.25);
\coordinate (A9) at (-4.5+3,-2.25);
\coordinate (A10) at (-4.5+3,-3);
\coordinate (A11) at (-5+3,-3);
\coordinate (A12) at (-5+3,-3.35);
\coordinate (A13) at (-5.5+3,-3.35);
\coordinate (A14) at (-5.5+3,-4);
\coordinate (A15) at (-6+3,-4);
\draw (A0) -- (A1) -- (A2) -- (A3) -- (A4) -- (A5) -- (A6) -- (A7)
-- (A8) -- (A9) -- (A10) -- (A11) -- (A12) -- (A13) -- (A14) -- (A15);
\coordinate (Atop) at (-4+3,0);
\node[above] at (Atop) {$Y_{I}$};

\coordinate (1stMajtop) at (-1+4,-0.8);
\node at (1stMajtop) {\Large{$\prec$}};

\coordinate (1stMajbot) at (-1+4,-2.5);
\node at (1stMajbot) {\Large{$\succ$}};

\coordinate (B0) at (5,0);
\coordinate (B1) at (4+5,0);
\coordinate (Bbottomleft) at (0+5,-4);
\draw (B0) -- (Bbottomleft);
\coordinate (B2) at (4+5,-0.55);
\coordinate (B3) at (-3+6+5,-0.55);
\coordinate (B4) at (-3+6+5,-1.05);
\coordinate (B5) at (-3.5+6+5,-1.05);
\coordinate (B6) at (-3.5+6+5,-1.8);
\coordinate (B7) at (-4+6+5,-1.8);
\coordinate (B8) at (-4+6+5,-2.15);
\coordinate (B9) at (-4.5+6+5,-2.15);
\coordinate (B10) at (-4.5+6+5,-3);
\coordinate (B11) at (-5+6+5,-3);
\coordinate (B12) at (-5+6+5,-3.3);
\coordinate (B13) at (-5.5+6+5,-3.3);
\coordinate (B14) at (-5.5+6+5,-4);
\coordinate (B15) at (-6+6+5,-4);
\draw (B0) -- (B1) -- (B2) -- (B3) -- (B4) -- (B5) -- (B6) -- (B7)
-- (B8) -- (B9) -- (B10) -- (B11) -- (B12) -- (B13) -- (B14) -- (B15);
\coordinate (Btop) at (2+5,0);
\node[above] at (Btop) {$Y_{F}$};

\coordinate (Y1Y2left) at (-6+3,-1.6);
\coordinate (Y1Y2right) at (2.45+5,-1.6);
\draw [thick , dashed] (Y1Y2left) -- (Y1Y2right);
\node[left] at (Y1Y2left) {$s$};

\draw[thick, ->] (0.3,-1.4) arc (-60:-10:1.2);
\draw[thick, ->] (-0.5,-2) arc (-10:-60:1.2);

\end{tikzpicture}
\caption{Figure showing schematically Young diagram $Y_{I}$ on the left and $Y_{F}$ on the right. Row $s$ in this pair is indicated by the dashed line. The arrows indicate a net movement of boxes upward and downward in diagram $Y_{I}$ to form $Y_{F}$.}
\label{fig1aa}
\end{figure}
\end{center}
\begin{definition}\label{partialmajdef}
	We define partial majorization with respect to row $s$, denoted by the symbol $\prec_{s}$. For the pair of Young diagrams $(Y,Y')$ of $n$ boxes, we say $Y'$ partially majorizes $Y$, written $Y \prec_{s} Y'$, if $Y \prec Y'$ above row $s$, and $Y \succ Y'$ for row $s$ and below.
\end{definition}
We give two simple examples to illustrate the above procedure. Consider $Y = (3,3)$. We perform two one-box moves to obtain $Y' = (4,1,1)$:
\bea
\label{eq:onebocmoveex1}
\ytableausetup{boxsize=1.2em}
\begin{ytableau}
       \none &&& \\
  \none  &&&
\end{ytableau} \hspace{20pt} \rightarrow \hspace{20pt}\ytableausetup{boxsize=1.2em}
\begin{ytableau}
       \none &&& \\
  \none  &&\\
  \none &
\end{ytableau}  \hspace{20pt} \rightarrow \hspace{20pt}\ytableausetup{boxsize=1.2em}
\begin{ytableau}
       \none &&&& \\
  \none  &\\
  \none &
\end{ytableau} 
\eea
Thus, $k^{(Y,Y')}_{swap} = 1$ with the swap occurring in row 2. We therefore write $(3,3) \prec_{2} (4,1,1)$. This example has already been mentioned in section \ref{domincomp} and is indeed a case for which $\Delta^{(Y',Y)}_{2} = 0$. As a final example, consider again $Y=(4,2,2,1)$. We again perform two one-box moves to obtain $Y' = (5,1,1,1,1)$:
\bea
\label{eq:onebocmoveex1}
\ytableausetup{boxsize=1.2em}
\begin{ytableau}
       \none &&&& \\
  \none  &&\\
  \none &&\\
  \none &
\end{ytableau} \hspace{20pt} \rightarrow \hspace{20pt}\ytableausetup{boxsize=1.2em}
\begin{ytableau}
       \none &&&& \\
  \none  &&\\
  \none &\\
  \none &\\
  \none &
\end{ytableau}  \hspace{20pt} \rightarrow \hspace{20pt}\ytableausetup{boxsize=1.2em}
\begin{ytableau}
       \none &&&&& \\
  \none  &\\
  \none &\\
  \none &\\
  \none &
\end{ytableau} 
\eea
$k^{(Y,Y')}_{swap} = 1$ with the swap ocurring in row 3. Thus, $(4,2,2,1) \prec_{3} (5,1,1,1,1)$.\\

Returning to the general discussion, $k^{(Y_{F},Y_{I})}_{swap} = 1$ with $Y_{I} \prec_{s} Y_{F}$. Further, assume $\Delta^{(Y_{F},Y_{I})}_{2} = 0$.  Then,
\bea
\label{eq:5.3a}
	\Delta^{(Y_{F},Y_{I})}_{2} &=& \sum\limits^{P-1}_{i=0}\Delta^{(Y_{i+1},Y_{i})}_{2} +  \sum\limits^{N-1}_{i=P+1}\Delta^{(Y_{i+1},Y_{i})}_{2} = 0, \\
	 Y_{0} &\equiv& Y_{I},  \; Y_{P} \equiv Y_{int} , \;  Y_{N}\equiv Y_{F}.\nonumber
\eea
Since $Y_{I} \prec Y_{F}$ for rows above $s$ and $Y_{I} \succ Y_{F}$ for rows below and including $s$, we have
\bea
\label{eq:5.3a1}
	\Delta^{(Y_{int},Y_{I})}_{2} &=& \sum\limits^{P-1}_{i=0}\Delta^{(Y_{i+1},Y_{i})}_{2} < 0,\\
\label{eq:5.3a2}
	\Delta^{(Y_{F},Y_{int+1})}_{2} &=& \sum\limits^{N-1}_{i=P+1}\Delta^{(Y_{i+1},Y_{i})}_{2} > 0.
\eea
We now study
\bea
\label{eq:5.3b}
	\Delta^{(Y_{F},Y_{I})}_{3} &=& \sum\limits^{P-1}_{i=0}\Delta^{(Y_{i+1},Y_{i})}_{3} + \sum\limits^{N-1}_{i=P+1}\Delta^{(Y_{i+1},Y_{i})}_{3}.
\eea
We now use the following result. From equation (\ref{eq:Sec5.1c}), one may show that if a single box is moved from row $\alpha$ of length $y_{i,\alpha}$ in $Y_{i}$ up to row $\beta$ of length $y_{i,\beta}$, where $\alpha>\beta$, to form $Y_{i+1}$, then
\bea
\label{eq:5.3c}
	\Delta^{(Y_{i+1},Y_{i})}_{3} = \frac{3}{2} p_{1}\left( \tilde{y}_{i,\alpha} , \tilde{y}_{i,\beta} \right)\Delta^{(Y_{i+1},Y_{i})}_{2} ,
\eea
where $p_{1}\left( \tilde{y}_{i,\alpha} , \tilde{y}_{i,\beta} \right)$ is the first power sum in the shifted row length variables $\tilde{y}_{i,\alpha}$ and $\tilde{y}_{i,\beta}$: $p_{1}\left( \tilde{y}_{i,\alpha} , \tilde{y}_{i,\beta} \right) = \tilde{y}_{i,\alpha} + \tilde{y}_{i,\beta}$. The same equation is true if the box is moved downward from row $\alpha$ to row $\beta$, where in this case $\alpha < \beta$. Equation (\ref{eq:5.3b}) now becomes
\bea
\label{eq:5.3d}
	\Delta^{(Y_{F},Y_{I})}_{3} &=& \sum\limits^{P-1}_{i=0} \frac{3}{2} p_{1}\left( \tilde{y}_{i,\delta} , \tilde{y}_{i,\gamma} \right)\Delta^{(Y_{i+1},Y_{i})}_{2} +  \sum\limits^{N-1}_{i=P+1}  \frac{3}{2} p_{1}\left( \tilde{y}_{i,\alpha} , \tilde{y}_{i,\beta} \right)\Delta^{(Y_{i+1},Y_{i})}_{2} .
\eea
In the first term of (\ref{eq:5.3d}), indices $\delta$ and $\gamma$ refer to rows below, or including, row $s$. We have $\delta > \gamma \geq s$ for a box moving downward from row $\gamma$ to $\delta$. In the second term, indices $\alpha$ and $\beta$ refer to rows above row $s$. If one of the indices, $\alpha$ say, does refer to row $s$, then row $\beta$ must be above row $s$ indicating that a box is moved from row $s$ to row $\beta$ above. So, $s \geq \alpha > \beta$. Recall that the shifted row lengths obey $\tilde{y}_{i,1} > \tilde{y}_{i,2} > \cdots > \tilde{y}_{i,l(Y_{i})} \geq 0 $. Thus, all of the power sums in the top region are larger than all of the power sums in the bottom region. This includes the minimum power sum in the top region, and maximum power sum in the bottom region:
\bea
\label{eq:5.3e}
p_{1}\left(\tilde{y}_{i,\alpha} , \tilde{y}_{i,\beta}\right)_{min} > p_{1}\left( \tilde{y}_{i,\delta} , \tilde{y}_{i,\gamma}\right)_{max}.
\eea
Then
\bea
\label{eq:5.3f}
	\Delta^{(Y_{F},Y_{I})}_{3} &>&  p_{1}\left( \tilde{y}_{i,\delta} , \tilde{y}_{i,\gamma} \right)_{max} \sum\limits^{P-1}_{i=0} \Delta^{(Y_{i+1},Y_{i})}_{2} +  p_{1}\left( \tilde{y}_{i,\alpha} , \tilde{y}_{i,\beta} \right)_{min} \sum\limits^{N-1}_{i=P+1}  \Delta^{(Y_{i+1},Y_{i})}_{2} ,\\
\label{eq:5.3g}
	&=& p_{1}\left( \tilde{y}_{i,\delta} , \tilde{y}_{i,\gamma} \right)_{max}  \Delta^{(Y_{int},Y_{I})}_{2}  +  p_{1}\left( \tilde{y}_{i,\alpha} , \tilde{y}_{i,\beta} \right)_{min}  \Delta^{(Y_{F},Y_{int+1})}_{2} .
\eea
Using equations (\ref{eq:5.3a}), (\ref{eq:5.3a1}) and (\ref{eq:5.3a2}), as well as (\ref{eq:5.3e}), we conclude that the right hand side of (\ref{eq:5.3g}) is positive. Thus,
\bea
\label{eq:5.3h}
	\Delta^{(Y_{F},Y_{I})}_{3} &>&0.
\eea
In this section, we have shown that if we have $k^{(Y_{F},Y_{I})}_{swap} = 1$ and $\Delta^{(Y_{F},Y_{I})}_{2} = 0$, then we must have $\Delta^{(Y_{F},Y_{I})}_{3} >0$. $Y_{I}$ and $Y_{F}$ were related to each other by beginning from $Y_{I}$ and performing a net downward movement of boxes from row $s$, and then a net upward movement of boxes from row $s$. For this set up, the normalized character of $T_{3}$ for $Y_{I}$ and $Y_{F}$ is guaranteed to be distinct. Thus, $k^{(Y_{I},Y_{F})}_{*} = 3$. Lastly, we draw the implication from this that if $\Delta^{(Y_{F},Y_{I})}_{2} = 0$ and $\Delta^{(Y_{F},Y_{I})}_{3} = 0$, then $k^{(Y_{I},Y_{F})}_{swap} = 1$ must be false. We thus need $k^{(Y_{F},Y_{I})}_{swap} \geq 2$.
\subsection{The limited $k^{(Y,Y')}_{swap} = 2$ case}\label{kswapeq2}

In this section, we assume that $Y_{I}$ and $Y_{F}$ have two swaps in the direction of the inequality sign in their row sum relations. We further assume that $\Delta^{(Y_{F},Y_{I})}_{2} = \Delta^{(Y_{F},Y_{I})}_{3} = 0$. This indicates that $\widehat{\chi}(T_{2})$ and $\widehat{\chi}(T_{3})$ are degenerate for these two Young diagrams. These Young diagrams may be divided into three regions. A top region where $Y_{I} \prec Y_{F}$, a middle region where $Y_{I} \succ Y_{F}$, and a bottom region where $Y_{I} \prec Y_{F}$ again. Definition \ref{partialmajdef} may naturally be extended to the case of two swaps. If the swaps occur in rows $s_{1}$ and $s_{2}$, with $s_{1} < s_{2}$, then $Y_{F}$ partially majorizes $Y_{I}$ with respect to $s_{1}$ and $s_{2}$, written $Y_{I} \prec_{s_{1}s_{2}} Y_{F}$, if $Y_{I} \prec Y_{F}$ for rows above $s_{1}$, and $Y_{I} \succ Y_{F}$ for rows $s_{1}$ down to row $s_{2}-1$, and $Y_{I} \prec Y_{F}$ for rows $s_{2}$ and below. In this section we only consider the simplest case where $Y_{I}$ and $Y_{F}$ are related by a movement of three boxes, the minimum needed for the necessary two swaps. 

We begin with partition $Y_{I}$
\bea
	Y_{I} = (\cdots , y_{I,1} , y_{I,2} , \cdots , y_{I,3} , y_{I,4} , \cdots , y_{I,5} , y_{I,6} , \cdots  ),
\eea
where $y_{I,1}$ and $y_{I,2} $ refer to two rows in the top region, $y_{I,3}$ and $y_{I,4}$ refers to two rows in the middle region, and $y_{I,5}$ and $y_{I,6}$ refer to any two rows in the bottom region. Any pair of these rows are not necessarily adjacent. We now perform three one-box moves to generate a sequence of diagrams $Y_{I} \rightarrow Y_{1} \rightarrow Y_{2} \rightarrow Y_{F} $, arriving at $Y_{F}$. Explicitly,
\bea
	Y_{I} &=& (y_{I,1} , y_{I,2} , \cdots , y_{I,3} , y_{I,4} , \cdots , y_{I,5} , y_{I,6} ), \nonumber \\
\label{eq:5.40}
	Y_{1} &=& (y_{I,1} + 1 , y_{I,2} - 1 , \cdots , y_{I,3} , y_{I,4} , \cdots , y_{I,5} , y_{I,6} ), \\
	Y_{2} &=& (y_{I,1} + 1 , y_{I,2} - 1 , \cdots , y_{I,3} - 1 , y_{I,4} + 1 , \cdots , y_{I,5} , y_{I,6} ), \nonumber\\
	Y_{F} &=& (y_{I,1} + 1 , y_{I,2} - 1 , \cdots , y_{I,3} - 1 , y_{I,4} + 1 , \cdots , y_{I,5} +1 , y_{I,6} - 1 ). \nonumber
\eea
This is schematically shown in Figure \ref{fig1ab}.
\begin{center}
\begin{figure}[h!]
\begin{tikzpicture}[semithick, >=Stealth]
\coordinate (Am1) at (-5,0);
\node[above] at (Am1) {.};
\coordinate (A0) at (0,0);
\coordinate (A1) at (4,0);
\coordinate (Abottomleft) at (0,-4);
\draw (A0) -- (Abottomleft);
\coordinate (A2) at (-2+6,-0.5);
\coordinate (A3) at (-3+6,-0.5);
\coordinate (A4) at (-3+6,-1);
\coordinate (A5) at (-3.5+6,-1);
\coordinate (A6) at (-3.5+6,-1.75);
\coordinate (A7) at (-4+6,-1.75);
\coordinate (A8) at (-4+6,-2.25);
\coordinate (A9) at (-4.5+6,-2.25);
\coordinate (A10) at (-4.5+6,-3);
\coordinate (A11) at (-5+6,-3);
\coordinate (A12) at (-5+6,-3.35);
\coordinate (A13) at (-5.5+6,-3.35);
\coordinate (A14) at (-5.5+6,-4);
\coordinate (A15) at (-6+6,-4);
\draw (A0) -- (A1) -- (A2) -- (A3) -- (A4) -- (A5) -- (A6) -- (A7)
-- (A8) -- (A9) -- (A10) -- (A11) -- (A12) -- (A13) -- (A14) -- (A15);
\coordinate (Atop) at (-4+6,0);
\node[above] at (Atop) {$Y_{I}\rightarrow Y_{F}$};

\coordinate (Y1Y2left) at (0,-1.4);
\coordinate (Y1Y2right) at (4,-1.4);
\draw [thick , dashed] (Y1Y2left) -- (Y1Y2right);
\coordinate (Y1Y2farright) at (4,-1.4);

\coordinate (Y1Y2leftbot) at (0,-2.5);
\coordinate (Y1Y2rightbot) at (4,-2.5);
\draw [thick , dashed] (Y1Y2leftbot) -- (Y1Y2rightbot);

\draw[thick, ->] (3.3,-1.3) arc (-60:-10:1);
\draw[draw=black] (4,-1.1) rectangle ++(0.15,0.15);

\draw[thick, ->] (3,-1.6) arc (-10:-60:1);
\draw[draw=black] (3.2,-2.1) rectangle ++(0.15,0.15);

\draw[thick, ->] (1.25,-3.75) arc (-60:-10:1);
\draw[draw=black] (2,-3.5) rectangle ++(0.15,0.15);

\end{tikzpicture}
\caption{Figure showing schematically our three box move to obtain Young diagram $Y_{F}$ from $Y_{I}$. Each arrow indicates a movement of a single box in either the top, middle, or bottom regions. }
\label{fig1ab}
\end{figure}
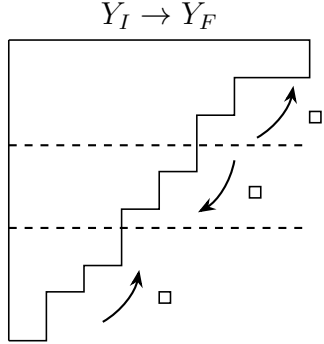
\end{center}
Thus $Y_{I}$ and $Y_{F}$ have two swaps in their row sum relations. We can again write the $\Delta$'s as a sum of contributions from each step in the path from $Y_{I}$ to $Y_{F}$
\bea
\label{eq:5.4a}
	\Delta^{(Y_{F},Y_{I})}_{2} = \Delta^{(Y_{1},Y_{I})}_{2} + \Delta^{(Y_{2},Y_{1})}_{2} + \Delta^{(Y_{F},Y_{2})}_{2} &=& 0,\\
\label{eq:5.4b}
	\Delta^{(Y_{F},Y_{I})}_{3} = \Delta^{(Y_{1},Y_{I})}_{3} + \Delta^{(Y_{2},Y_{1})}_{3} + \Delta^{(Y_{F},Y_{2})}_{3} &=& 0,\\
\label{eq:5.4c}
	\Delta^{(Y_{F},Y_{I})}_{4} = \Delta^{(Y_{1},Y_{I})}_{4} + \Delta^{(Y_{2},Y_{1})}_{4} + \Delta^{(Y_{F},Y_{2})}_{4}.
\eea
We present a detailed calculation in Appendix \ref{2swap} showing that $\Delta^{(Y_{F},Y_{I})}_{4} > 0 $. We again make use of result (\ref{eq:5.3c}) and of an analogous result for $\Delta^{(Y_{i+1},Y_{i})}_{4}$:
\bea
\label{eq:5.4d}
	\Delta^{(Y_{i+1},Y_{i})}_{4} = \frac{2}{3} p_{1}\left( \tilde{y}_{i,\alpha} , \tilde{y}_{i,\beta} \right)\Delta^{(Y_{i+1},Y_{i})}_{3} +  p_{2}\left( \tilde{y}_{i,\alpha} , \tilde{y}_{i,\beta} \right)\Delta^{(Y_{i+1},Y_{i})}_{2} + \frac{1}{2}\left( \Delta^{(Y_{i+1},Y_{i})}_{2} \right)^{2},
\eea
where a single box is moved from some row $\alpha$ up to some row $\beta$, $\alpha > \beta$. The same expression is also true for a single box downward move. We have derived similar expressions for some higher $\Delta_{k}$ and list them in Appendix \ref{DeltaID}. Thus, for diagrams $Y_{I}$ and $Y_{F}$ related by the three box move described above, giving $k^{(Y_{I},Y_{F})}_{swap} = 2$, and with  $\Delta^{(Y_{F},Y_{I})}_{2} = \Delta^{(Y_{F},Y_{I})}_{3} = 0$, we have $\Delta^{(Y_{F},Y_{I})}_{4} > 0 $. This means that $\widehat{\chi}(T_{4})$ is distinct for these diagrams. This implies that if $\Delta^{(Y_{F},Y_{I})}_{2} = \Delta^{(Y_{F},Y_{I})}_{3} = 0$ and $\Delta^{(Y_{F},Y_{I})}_{4} = 0$, then $k^{(Y_{F},Y_{I})}_{swap} = 2$ must be false. Thus, $k^{(Y_{F},Y_{I})}_{swap} \geq 3$.

In the last part of this section, we describe an extension of this result. Let two diagrams $Y_{1}$ and $Y_{2}$ be related by the three-box move described in (\ref{eq:5.40}). Consequently $k^{(Y_{2},Y_{1})}_{swap} = 2$. Next, let $Y_{2}$ and $Y_{3}$ also be related by a similar three box move making $k^{(Y_{3},Y_{2})}_{swap} = 2$. The rows in which boxes are moved to obtain $Y_{3}$ from $Y_{2}$ need not be the same as those involved for $Y_{1}$ and $Y_{2}$. Then, $Y_{1}$ and $Y_{3}$ will also be incomparable and have at least two swaps of the inequality sign in their row sum relations -  $k^{(Y_{1},Y_{3})}_{swap} \geq 2$ (see discussion below and Appendix \ref{Transitivity}). We again assume that $\Delta^{(Y_{2},Y_{1})}_{2} = \Delta^{(Y_{2},Y_{1})}_{3} = 0$, and also that $\Delta^{(Y_{3},Y_{2})}_{2} = \Delta^{(Y_{3},Y_{2})}_{3} = 0$. From our discussion above, $\Delta^{(Y_{2},Y_{1})}_{4}>0$ and $\Delta^{(Y_{3},Y_{2})}_{4}>0$, and since 
\bea
\label{eq:5.4e}
	\Delta^{(Y_{3},Y_{1})}_{p} = \Delta^{(Y_{2},Y_{1})}_{p}  + \Delta^{(Y_{3},Y_{2})}_{p} , \hspace{20pt} p \geq 1
\eea
we have $\Delta^{(Y_{3},Y_{1})}_{4} > 0 $. This can be extended to the sequence $Y_{1} \rightarrow Y_{2} \rightarrow \cdots \rightarrow Y_{n}$, where each $\rightarrow$ means the three box move (\ref{eq:5.40}). Further $k^{(Y_{i+1},Y_{i})}_{swap} = 2 $ and thus $k^{(Y_{n},Y_{1})}_{swap} \geq 2$. If, at each step along the way, we have $\Delta^{(Y_{i+1},Y_{i})}_{2} = \Delta^{(Y_{i+1},Y_{i})}_{3} = 0$, then $\Delta^{(Y_{i+1},Y_{i})}_{4} > 0$. Using the fact that
\bea
	\Delta^{(Y_{n},Y_{1})}_{p} = \sum\limits^{n-1}_{i=1} \Delta^{(Y_{i+1},Y_{i})}_{p},
\eea
we are guaranteed that $\Delta^{(Y_{n},Y_{1})}_{4} > 0$. Even though each $Y_{i}$ and $Y_{i+1}$ differ by our simple three box move described in (\ref{eq:5.40}), the pair $Y_{1}$ and $Y_{n}$ have at least two swaps, and may differ quite substantially from each other. Of course, the assumption that $\Delta^{(Y_{i+1},Y_{i})}_{p} = 0$, for $p=2,3$ for each step in the sequence is not true in general. Nonetheless, this result is consistent with our conjecture (\ref{eq:2c}).

In general, it is known that Young diagrams and their majorization/dominance ordering possesses the transitivity property. That is, for any three diagrams $Y_{1}, Y_{2}$ and $Y_{3}$ such that $Y_{1} \prec Y_{2}$, and $Y_{2} \prec Y_{3}$, we have $Y_{1} \prec Y_{3}$. However this property has a generalization to incomparable Young diagrams for our measure of incomparability, $k_{swap}$. For any three diagrams $Y_{1}, Y_{2}$ and $Y_{3}$ such that $k^{(Y_{2},Y_{1})}_{swap} = 1$, and $k^{(Y_{3},Y_{2})}_{swap} = 1$, we have $k^{(Y_{3},Y_{1})}_{swap} \geq 1$. Further, for any three diagrams $Y_{1}, Y_{2}$ and $Y_{3}$ such that $k^{(Y_{2},Y_{1})}_{swap} = 2$, and $k^{(Y_{3},Y_{2})}_{swap} = 2$, we have $k^{(Y_{3},Y_{1})}_{swap} \geq 2$. We present some heuristic arguments supporting this in Appendix \ref{Transitivity}. 

\section{Determining the smallest $n$ for increasing incomparability}\label{Asymp}

According to our conjecture, degeneracies in the normalized characters of the cycle operators $\{T_{2} , T_{3} , \cdots \}$ amongst the irreps of $S_{n}$ requires a minimum number of swaps in the inequalities in the row sum relations of the corresponding Young diagrams. Recall that the integer $k^{(n)}_{*}$ is the minimum number of cycle operators needed to distinguish all the $S_{n}$ irreps. A corollary of our conjecture states that the maximum number of swaps in row sum relations between all pairs of these irreps acts as an upper bound for $k^{(n)}_{*}$. 
As an avenue to study $k^{(n)}_{(swap,max)}$, we propose an algorithm to construct a Young diagram pair having $l$ swaps using the least number of boxes $n$ possible. We call this $n$ value $n^{(l)}_{min}$ and is the minimum $n$ for which Young diagram pairs appear having $l$ swaps in their row sum relations. This implies that for $n = n^{(l)}_{min}$, $k^{(n)}_{(swap,max)} = l$ which remains the maximum number of swaps amongst Young diagram pairs up until some $n^{(l)}_{max}$. The integer $n^{(l)}_{max}$ is the maximum $n$ for which $l$ is the maximum number of swaps in the row sum relations between Young diagram pairs. If $ n$ increases any further, then there exist pairs having a maximum of $l+1$ swaps. Thus, $n^{(l+1)}_{min} = n^{(l)}_{max} + 1$. From the construction described below, we can obtain $n^{(l)}_{min}$ as a function of $l$. 


\subsection{General construction}

Begin constructing the pair $(Y,Y')$ with $Y = (0)$ and $Y' = (1)$. We write the partitions in non-decreasing order. The symbol of $(Y,Y')$ is simply $(<)$. Now we add integers to $Y$ and $Y'$ in the way described by the conjecture below.\\
\begin{conjecture}\label{Conj2}
To construct a pair of Young diagrams for the minimum value of $n$ with $l$ swaps in the row sums relations, the optimal symbol for the pair is
\bea
\label{eq:7m}
	(<,=,>,=,<,=, > , \cdots , >,<,>, < , \cdots , =).
\eea
\end{conjecture}
When adding an integer to $Y$ and $Y'$, it is the smallest integer possible so that we generate a swap, or an equals sign, and so that $Y$ and $Y'$ remain valid Young diagrams. Initially in (\ref{eq:7m}) the sequence alternates between $=$ signs and swaps. At some point in the sequence, we no longer have $=$ signs and the remaining entries are only swaps. How many $=$ signs should we have? If $l$ is even, there are $l/2$ number of $=$ signs in (\ref{eq:7m}), and if $l$ is odd we can have $\lfloor l/2 \rfloor$ or $\lceil l/2 \rceil$ number of $=$ signs. We do not count the final $=$ sign in this construction. After we have generated the required number of $=$ signs, we add the integers, in accordance with the above condition, that will generate the remaining number of swaps. Once we have constructed a pair of Young diagrams, summing their parts gives $n^{(l)}_{min}$. We describe this procedure in detail for $l=1,2,3$ in Appendix \ref{ConstructPart}.

We give here the example of $l = 5$. $l$ is odd, so we construct a pair $(Y_{1},Y'_{1})$ with 3 $=$ signs in their row sum relations, and a pair $(Y_{2},Y'_{2})$ with 2 $=$ signs in their row sum relations. Explicitly, we construct pairs whose symbols are 
\bea
	\{Y_{1} , Y'_{1}\} \rightarrow (<,=,>,=,<,=,>,<,>,=), \\
	\{Y_{2} , Y'_{2}\} \rightarrow (<,=,>,=,<,>,<,>,=). \nonumber
\eea
At each step in the construction we add the smallest integer to generate either a swap or an $=$ sign while ensuring the result is still a valid Young diagram. We find
\bea
\label{eq:6i}
	Y_{1} &=& (0,2,2,2,2,4,4,4,8,8),\\
\label{eq:6j}
	Y'_{1} &=& (1,1,1,3,3,3,3,6,6,9), \nonumber\\
\label{eq:6k}
	Y_{2} &=& (0,2,2,2,2,5,5,9,9),\\
\label{eq:6l}
	Y'_{2} &=& (1,1,1,3,3,3,7,7,10), \nonumber
\eea
From these pairs, we see that $n^{(5)}_{min}=36$. Using this conjecture, we compute $n^{(l)}_{min}$ for up to partitions having $l=12$ swaps. See Table \ref{Tab:Tabnmin}. 

In order to check this conjecture, in Mathematica we generated all partitions at a given $n$, paired them up in all possible ways and the measured $k_{swap}$ for each pair. This allowed us to determine the actual values of $n^{(l)}_{min}$, i.e., the smallest $n$ for which $k^{(n)}_{(swap,max)} = l$. We generated data up to $n=36$. The $n^{(l)}_{min}$ values are in perfect agreement with Table \ref{Tab:Tabnmin}, giving some evidence for the conjecture. The partitions constructed using the conjecture for $l=4, 5 , \cdots , 11$ are listed in Appendix \ref{ConstructPart}.

\begin{table}[htp]
\begin{center}
\begin{tabular}{|c|c|}
\hline
 Number of swaps, $l$ & $n^{(l)}_{min}$ \\
 \hline
 1 & 6 \\
 \hline
 2 & 11 \\
 \hline
 3 & 18 \\
  \hline
 4 & 26 \\
  \hline
 5 & 36 \\
  \hline
 6 & 47 \\
  \hline
 7 & 60 \\
  \hline
 8 & 74 \\
  \hline
 9 & 90 \\
 \hline
 10 & 107 \\
 \hline
 11 & 126 \\
 \hline
  12 & 146 \\
 \hline
\end{tabular}
\caption{\small{Table showing the smallest $n$ value for which pairs of Young diagram pairs having a given number swaps first appear. This data has been generated using our conjecture. We have verified this data up to $l=5$ and $n=36$.}}
\label{Tab:Tabnmin}
\end{center}
\end{table}

Using the information in Table \ref{Tab:Tabnmin} we can check our conjecture for an upper bound for $k^{(n)}_{*}$ in (\ref{eq:2d}) by comparing with known values of $k^{(n)}_{*}$ listed in \cite{KempRam}, \cite{GelounRam1}. This comparison is displayed in Table \ref{Tab:Tabcompeverything}.
\begin{table}[htp]
\begin{center}
\begin{tabular}{|c|c|c|}
\hline
 $k^{(n)}_{*}$ (known) & $k^{(n)}_{swap,max} + 2$ & Range of $n$ \\
 \hline
 2 & 2 & $n = \{1, \cdots , 5\}$  \\
  \hline
  $\{2,3\}$ & 3 & $n = \{6, \cdots , 10\} $ \\
 \hline
  $\{3,4\}$ & 4 & $n = \{11, \cdots , 17\} $ \\
  \hline
    $\{3,4\}$ & 5 & $n = \{18, \cdots , 25\} $ \\
  \hline
   $\{4,5\}$ & 6 & $n = \{26, \cdots , 35\} $ \\
  \hline
  $\{5,6\}$ & 7 & $n = \{36, \cdots , 46\} $ \\
  \hline
   $6$ & 8 & $n = \{47, \cdots , 59\} $ \\
  \hline
   $6$ & 9 & $n = \{60, \cdots , 73\} $ \\
  \hline
  $\{6,7\}$ & 10 & $n = \{74, \cdots , 81\} $ \\
  \hline
    unknown & 10 & $n = \{82, \cdots , 89\} $ \\
  \hline
     unknown & 11 & $n = \{90, \cdots , 106\} $ \\
  \hline
    unknown & 12 & $n = \{107, \cdots , 125\} $ \\
  \hline
\end{tabular}
\caption{\small{Table comparing $k^{(n)}_{swap,max} + 2$ with actual known values of $k^{(n)}_{*}$ within certain ranges of $n$. At $n=80$, we know $k^{(n)}_{*} = 7$, but at $n=81$, $k^{(n)}_{*} = 6$. Values for $k^{(n)}_{*}$ above $n=81$ are unknown.}}
\label{Tab:Tabcompeverything}
\end{center}
\end{table}

Partly inspired by the analysis in section \ref{AsympBehav}, and the result in equation (\ref{eq:6n}), we find the following closed formula reproducing all the integers in Table \ref{Tab:Tabnmin}
\bea
\label{eq:closedformforn}
	n^{(l)}_{min} = \lceil \frac{3}{4}l^{2} + 3l + 2 \rceil, \hspace{20pt } l \geq 1 . 
\eea
It is fascinating that the sequence of integers in Table \ref{Tab:Tabnmin} has already been discovered and exists in the Online Encyclopedia of Integer Sequences \cite{OEIS}. This entry in \cite{OEIS} gives a closed formula for $a(n)$ that reproduces the sequence for $n^{(l)}_{min}$ for $n \geq 3$, 
\bea
\label{eq:closedformfornoeis}
	a(n) = \frac{1}{8}\left( -7 + (-1)^{1+n} + 6n^{2} \right), \hspace{20pt} n \geq 0.
\eea
We have confirmed that (\ref{eq:closedformforn}) is equal to (\ref{eq:closedformfornoeis}) for both cases when $l$ is even and when $l$ is odd. It is not clear to us, however, the context in which the sequence of integers in \cite{OEIS} was discovered. We are currently following up with OEIS.

\subsection{Asymptotic behaviour of $k^{(n)}_{(swap,max)}$}\label{AsympBehav}

\begin{corollary}
	At large $n$ and at large $l$, $k^{(n)}_{*}$ is bounded above according to
\bea
	k^{(n)}_{*} \leq k^{(n)}_{swap,max} \sim \frac{2}{\sqrt{3}} ( n )^{1/2}.\nonumber
\eea
\end{corollary}
To argue in favour of this corollary assume for now that $l$ and $l/2$ are both even. Then, one of the Young diagram pairs constructed according to the above algorithm is  
\bea
	(0,2,2,2,2,4,4,4,4,\cdots,\frac{l}{2},\frac{l}{2},\frac{l}{2},\frac{l}{2},\frac{l}{2}+3,\frac{l}{2}+3,\frac{l}{2}+7,\frac{l}{2}+7,\cdots , \frac{l}{2}+ (l-1), \frac{l}{2} +( l-1),\frac{3l}{2}+2 )\nonumber\\
\eea
To find the number of boxes in this diagram, we add all the parts. This will give $n^{(l)}_{min}$. This is done by
\bea
	4\sum\limits^{l/4}_{i=1} 2i &=& \frac{l}{4}(l+4).\\
	2\sum\limits^{l/4-1}_{i=0} \left( \frac{l}{2} + 3 +4i \right) &=& \frac{l}{2}(l+1).\\
	\frac{l}{2} + 3 +4 (\frac{l}{4}-1) + 3 &=& \frac{3l}{2}+2.
\eea
Putting all these together,
\bea
\label{eq:6n}
	n^{(l)}_{min} = \frac{l}{4}(l+4) + \frac{l}{2}(l+1) + \frac{3l}{2}+2 &=& \frac{3}{4} l^{2} + 3 l + 2.
\eea
Recall that $k^{(n)}_{(swap,max)} $ is equal to the maximum number of swaps in the row sum relations between all Young diagram pairs for a given $n$ and it is equal to some particular integer typically for a range of $n$ values,
\bea
\label{eq:6o}
	k^{(n)}_{(swap,max)} = l , \hspace{20pt} n \in \{ n^{(l)}_{min} , \cdots , n^{(l)}_{max} \}.
\eea
For large $l \gg 1$, from (\ref{eq:6n}) we see $n^{(l)}_{min} \sim 3l^{2} / 4$. Further, it is reasonable to assume that, since $l \gg 1$, we also have $n \gg 1$. Then $l \pm 1 \approx l$, and, since $n^{(l)}_{max} = n^{(l+1)}_{min} -1 $, we have $n^{(l)}_{max} \approx  n^{(l)}_{min} $. Thus, all $n$ values in the range $\{n_{min} , \cdots , n_{max}\}$ in (\ref{eq:6o}) scale like $3l^{2}/4$. Invert this relation to find $l \sim 2/\sqrt{3}\, (n)^{1/2}$. Using this we can express $k^{(n)}_{(swap,max)}$ in terms of $n$:
 \bea
 \label{eq:6p}
	k^{(n)}_{(swap,max)} \sim \frac{2}{\sqrt{3}}  n^{1/2}.
\eea
This behaviour is true for all $n$ values for which the maximum number of swaps in Young diagram row sums is approximately $l$. This is also true for $l\pm a$, where $a$ is $O(1)$, or even $O(\sqrt{l})$. Lastly, we conclude that at large $n$, where $l$ is also large, $k^{(n)}_{*}$ is bounded above according to
\bea
 \label{eq:6q}
	k^{(n)}_{*} \leq k^{(n)}_{swap,max} \sim \frac{2}{\sqrt{3}} n^{1/2}.
\eea
We have analysed the asymptotics of the upper bound for $k^{(n)}_{*}$. The upper bound was based on the notion of measuring the degree of incomparability between Young diagrams of $n$ boxes. The integer $k^{(Y,Y')}_{swap}$ counted the number of swaps in the direction of the inequality sign in the row sum relations of $Y$ and $Y'$ and played the role of our measure of the incomparability of the pair. The upper bound for $k^{(n)}_{*}$ involved the maximum number of swaps between all pairs of Young diagrams at fixed $n$, and so $k^{(n)}_{*}$ was bounded above by the maximum incomparability between the symmetric group $S_{n}$ irreps. A heuristic argument is given in \cite{GelounRam1} which finds $k^{(n)}_{*} \sim n^{1/4}$. Furthermore, their quantum detection algorithm was subject to the assumption that $k^{(n)}_{*} \in O(n^{\alpha}) , \alpha < 1/2$. We are finding in this work that, according to our conjectures and constructions, $k^{(n)}_{*}$ is indeed bounded by a function that scales as $n^{1/2}$.
 
\section{Discussion}

The theory of majorization and the dominance ordering has offered some insight into the problem of distinguishing irreps of the symmetric group $S_{n}$. We relied on the list of normalized characters of elements of $\mathcal{Z}[\mathbb{C}(S_{n})]$ to distinguish between the irreps. The central elements were the cycle operators $T_{k}$, and were labeled by conjugacy classes of permutations corresponding to a single permutation of length $k$. The dominance ordering, and the incomparability measure we introduced played a key role in the conjecture posited in this work. This conjecture expressed a necessary condition for the irreps $R$ of $S_{n}$ to have degenerate $\{ \widehat{\chi}_{R}(T_{2}) , \cdots , \widehat{\chi}_{R}(T_{k}) \}$. When there are degeneracies we are not able to distinguish between all the symmetric group $S_{n}$ irreps and the subset $\{T_{2} ,\cdots , T_{k} \}$ does not generate $\mathcal{Z}[\mathbb{C}(S_{n})]$. The minimum number of cycle operators needed to generate the centre and thus to distinguish all $S_{n}$ irreps was the integer $k^{(n)}_{*}$. Through Schur-Weyl duality, $k^{(n)}_{*}$ was the minimum number of Casimir operators in the $U(N)$ gauge theory needed to uniquely identify the 1/2-BPS state of energy  $n$. Through AdS/CFT, this is related to the minimum number of multipole moments an observer needs to measure to identify the dual state. It is known that when the energy $n$ of the state scales as $O(N^{2})$, the AdS states are new geometries which includes extremal black holes. For an observer in AdS looking to distinguish between these bulk geometries, or so-called heavy states, it is interesting to ask up to what energy scale do they need to measure multipole moments to? In the semi-classical limit, an observer in AdS can only measure these quantities up to the Planck scale which scales as $N^{1/4}$. In our work considerations of incomparability amongst 1/2-BPS states lead to a derivation of an upper bound for $k^{(n)}_{*}$ that scaled as $n^{1/2}$. The precise asymptotic behaviour of $k^{(n)}_{*}$ may lead to interesting insights regarding the energy scales an observer in the full theory (beyond the semi-classical regime) is required to measure up to in order to uniquely distinguish between the black hole like states. We plan to study this further in the future.


\begin{center}
	\textbf{Acknowledgments}
\end{center}
The author would like to thank Sanjaye Ramgoolam for collaboration in the early stages of this project and discussions on the formulation of the conjecture. The author would also like to thank Amartya Goswarmi for helpful and pleasant discussions. 

\appendix

\section{Explicit two swap calculation}\label{2swap}

Using (\ref{eq:5.3c}) and then condition (\ref{eq:5.4a}), equation (\ref{eq:5.4b}) becomes
\bea
	\Delta^{(Y_{F},Y_{I})}_{3} &=& \frac{3}{2}p_{1}( \tilde{y}_{0,1} , \tilde{y}_{0,2} )\Delta^{(Y_{1},Y_{I})}_{2} + \frac{3}{2}p_{1}( \tilde{y}_{0,3} , \tilde{y}_{0,4} )\Delta^{(Y_{2},Y_{1})}_{2} + \frac{3}{2}p_{1}( \tilde{y}_{0,5} , \tilde{y}_{0,6} )\Delta^{(Y_{F},Y_{2})}_{2} = 0,\nonumber \\
	&=& \left[ p_{1}( \tilde{y}_{0,1} , \tilde{y}_{0,2} ) -  p_{1}( \tilde{y}_{0,3} , \tilde{y}_{0,4} ) \right] \Delta^{(Y_{1},Y_{I})}_{2} - \left[ p_{1}( \tilde{y}_{0,3} , \tilde{y}_{0,4} ) -  p_{1}( \tilde{y}_{0,5} , \tilde{y}_{0,6} ) \right] \Delta^{(Y_{F},Y_{2})}_{2} = 0. \nonumber \\
\label{eq:Aa}
\eea
We now use the following result. If a single box is moved from row $\alpha$ of length $y_{i,\alpha}$ in $Y_{i}$ up to row $\beta$ of length $y_{i,\beta}$, where $\alpha>\beta$, to form $Y_{i+1}$, then
\bea
\label{eq:Ab}
	\Delta^{(Y_{i+1},Y_{i})}_{4} = \frac{2}{3} p_{1}\left( \tilde{y}_{i,\alpha} , \tilde{y}_{i,\beta} \right)\Delta^{(Y_{i+1},Y_{i})}_{3} +  p_{2}\left( \tilde{y}_{i,\alpha} , \tilde{y}_{i,\beta} \right)\Delta^{(Y_{i+1},Y_{i})}_{2} + \frac{1}{2}\left( \Delta^{(Y_{i+1},Y_{i})}_{2} \right)^{2},
\eea
where $p_{2}$ is the second order power sum, $p_{2}(x,y) = x^{2} + y^{2}$. Thus, $\Delta^{(Y_{F},Y_{I})}_{4}$ is
\bea
	\Delta^{(Y_{F},Y_{I})}_{4} &=& \frac{2}{3} p_{1}\left( \tilde{y}_{0,1} , \tilde{y}_{0,2} \right)\Delta^{(Y_{1},Y_{I})}_{3} +  p_{2}\left( \tilde{y}_{0,1} , \tilde{y}_{0,2} \right)\Delta^{(Y_{1},Y_{I})}_{2} + \frac{1}{2}\left( \Delta^{(Y_{1},Y_{I})}_{2} \right)^{2} \nonumber \\
\label{eq:Ac}
	&+&\frac{2}{3} p_{1}\left( \tilde{y}_{0,3} , \tilde{y}_{0,4} \right)\Delta^{(Y_{2},Y_{1})}_{3} +  p_{2}\left( \tilde{y}_{0,3} , \tilde{y}_{0,4} \right)\Delta^{(Y_{2},Y_{1})}_{2} + \frac{1}{2}\left( \Delta^{(Y_{2},Y_{1})}_{2} \right)^{2}\\
	&+& \frac{2}{3} p_{1}\left( \tilde{y}_{0,5} , \tilde{y}_{0,6} \right)\Delta^{(Y_{3},Y_{2})}_{3} +  p_{2}\left( \tilde{y}_{0,5} , \tilde{y}_{0,6} \right)\Delta^{(Y_{3},Y_{2})}_{2} + \frac{1}{2}\left( \Delta^{(Y_{3},Y_{2})}_{2} \right)^{2}. \nonumber
\eea
Study the $\Delta_{3}$ terms. Again using (\ref{eq:5.3c}) with (\ref{eq:5.4a}) for these terms,
\bea
	&& p_{1}(\tilde{y}_{0,1},\tilde{y}_{0,2})^{2}\Delta^{(Y_{1},Y_{I})}_{2} + p_{1}(\tilde{y}_{0,3},\tilde{y}_{0,4})^{2}\Delta^{(Y_{2},Y_{1})}_{2} + p_{1}(\tilde{y}_{0,5},\tilde{y}_{0,6})^{2}\Delta^{(Y_{3},Y_{2})}_{2} \nonumber \\
	&=& \left[ p_{1}(\tilde{y}_{0,1},\tilde{y}_{0,2})^{2} - p_{1}(\tilde{y}_{0,3},\tilde{y}_{0,4})^{2}\right] \Delta^{(Y_{1},Y_{I})}_{2}  - \left[p_{1}(\tilde{y}_{0,3},\tilde{y}_{0,4})^{2} - p_{1}(\tilde{y}_{0,5},\tilde{y}_{0,6})^{2} \right]\Delta^{(Y_{3},Y_{2})}_{2} \nonumber\\
	&=& \left[ p_{1}(\tilde{y}_{0,1},\tilde{y}_{0,2}) +   p_{1}(\tilde{y}_{0,3},\tilde{y}_{0,4})\right] \left[ p_{1}(\tilde{y}_{0,1},\tilde{y}_{0,2})  -  p_{1}(\tilde{y}_{0,3},\tilde{y}_{0,4})\right]  \Delta^{(Y_{1},Y_{I})}_{2} \nonumber \\
	&& - \left[ p_{1}(\tilde{y}_{0,3},\tilde{y}_{0,4}) +  p_{1}(\tilde{y}_{0,5},\tilde{y}_{0,6})\right] \left[ p_{1}(\tilde{y}_{0,3},\tilde{y}_{0,4})  -  p_{1}(\tilde{y}_{0,5},\tilde{y}_{0,6})\right]  \Delta^{(Y_{F},Y_{2})}_{2} \nonumber \\
	&=&  \left[ p_{1}(\tilde{y}_{0,1},\tilde{y}_{0,2}) +   p_{1}(\tilde{y}_{0,3},\tilde{y}_{0,4}) - p_{1}(\tilde{y}_{0,3},\tilde{y}_{0,4}) -  p_{1}(\tilde{y}_{0,5},\tilde{y}_{0,6}) \right] \left[ p_{1}(\tilde{y}_{0,1},\tilde{y}_{0,2})  -  p_{1}(\tilde{y}_{0,3},\tilde{y}_{0,4})\right]  \Delta^{(Y_{1},Y_{I})}_{2}  \nonumber 
 \eea
where we used (\ref{eq:Aa}) in the last line above. After canceling the $ p_{1}(\tilde{y}_{0,3},\tilde{y}_{0,4})$ terms, these terms become
\bea
	 \left[ p_{1}(\tilde{y}_{0,1},\tilde{y}_{0,2}) -  p_{1}(\tilde{y}_{0,5},\tilde{y}_{0,6}) \right] \left[ p_{1}(\tilde{y}_{0,1},\tilde{y}_{0,2})  -  p_{1}(\tilde{y}_{0,3},\tilde{y}_{0,4})\right]  \Delta^{(Y_{1},Y_{I})}_{2}.
\eea
Each factor in the above line is positive making the $\Delta_{3}$ terms in $\Delta^{(Y_{F},Y_{I})}_{4}$ is positive. We now focus on the $\Delta_{2}$ terms in $\Delta^{(Y_{F},Y_{I})}_{4}$. Using (\ref{eq:5.4a}), these terms are
\bea
	&&\left( \tilde{y}^{2}_{0,1} + \tilde{y}^{2}_{0,2} \right)\Delta^{(Y_{1},Y_{I})}_{2} + \left( \tilde{y}^{2}_{0,3} + \tilde{y}^{2}_{0,4} \right)\Delta^{(Y_{2},Y_{1})}_{2} + \left( \tilde{y}^{2}_{0,5} + \tilde{y}^{2}_{0,6} \right)\Delta^{(Y_{F},Y_{2})}_{2} \nonumber\\
		&=&\left( \tilde{y}^{2}_{0,1} + \tilde{y}^{2}_{0,2} - \tilde{y}^{2}_{0,3} - \tilde{y}^{2}_{0,4}  \right)\Delta^{(Y_{I},Y_{1})}_{2} - \left( \tilde{y}^{2}_{0,3} + \tilde{y}^{2}_{0,4} - \tilde{y}^{2}_{0,5} - \tilde{y}^{2}_{0,6} \right)\Delta^{(Y_{F},Y_{2})}_{2} \nonumber\\
		&=& \left[ \left( \tilde{y}_{0,1} + \tilde{y}_{0,3}\right) \left( \tilde{y}_{0,1} - \tilde{y}_{0,3}\right) + \left( \tilde{y}_{0,2} + \tilde{y}_{0,4}\right) \left( \tilde{y}_{0,2} - \tilde{y}_{0,4}\right)\right]\Delta^{(Y_{1},Y_{I})}_{2} \nonumber \\
		&& - \left[ \left( \tilde{y}_{0,3} + \tilde{y}_{0,5}\right) \left( \tilde{y}_{0,3} - \tilde{y}_{0,5}\right) + \left( \tilde{y}_{0,4} + \tilde{y}_{0,6}\right) \left( \tilde{y}_{0,4} - \tilde{y}_{0,6}\right)\right]\Delta^{(Y_{F},Y_{2})}_{2} \nonumber \\
		&>& \left( \tilde{y}_{0,2} + \tilde{y}_{0,4}\right) \left[  \tilde{y}_{0,1} - \tilde{y}_{0,3} + \tilde{y}_{0,2} - \tilde{y}_{0,4} \right ] \Delta^{(Y_{1},Y_{I})}_{2} - \left( \tilde{y}_{0,3} + \tilde{y}_{0,5} \right)\left[ \tilde{y}_{0,3} - \tilde{y}_{0,5} + \tilde{y}_{0,4} - \tilde{y}_{0,6} \right] \Delta^{(Y_{F},Y_{2})}_{2} \nonumber \\
		&=& \left( \tilde{y}_{0,2} + \tilde{y}_{0,4} -  \tilde{y}_{0,3} - \tilde{y}_{0,5} \right) \left[  \tilde{y}_{0,1} - \tilde{y}_{0,3} + \tilde{y}_{0,2} - \tilde{y}_{0,4} \right ] \Delta^{(Y_{1},Y_{I})}_{2},
\eea
where we again used (\ref{eq:Aa}) in the last line above. Each factor is again positive making these contributions to $\Delta^{(Y_{F},Y_{I})}_{4}$ positive. Lastly, the squared terms in (\ref{eq:Ac}) are positive. We thus conclude that $\Delta^{(Y_{F},Y_{I})}_{4} > 0$. 

There are other three box moves that produce two swaps. For example we could have, $\tilde{y}_{0,2} = \tilde{y}_{0,3}$. I.e. the same row can give one box upward and one box below. The result $\Delta^{(Y_{F},Y_{I})}_{4} > 0$ still applies in these cases with the analysis being very similar. Even though we present the simplest case it is plausible that it may provide a starting point for a general proof.

\section{Identities for $\Delta_{k}$ for small $k$}\label{DeltaID}

 Consider any two rows in $X$ and $Y$. In $X$, let there be $r$ 1-box moves from row $\alpha$ up to row $\beta$ to form $Y$:
\bea
	y_{\beta} &=& x_{\beta} + r \nonumber \\
	y_{\alpha} &=& x_{\alpha} - r \\
	\Delta^{(r_{\uparrow})}_{2} &=& h_{1}(y_{i},x_{i})(y_{i} - x_{i}) +h_{1}(y_{j},x_{j})(y_{j} - x_{j})\nonumber \\
	&=& 2r\left[ x_{i} - (x_{j} - r) \right].\\
	\Delta^{(r_{\uparrow})}_{3} &=& h_{2}(y_{i},x_{i})(y_{i} - x_{i}) +h_{2}(y_{j},x_{j})(y_{j} - x_{j}) \\
	\Delta^{(r_{\uparrow})}_{3} &=& \frac{1}{2}\left( p_{1}(y_{i},x_{i})h_{1}(y_{i},x_{i}) + p_{2}(y_{i},x_{i}) \right) (y_{i} - x_{i}) + \frac{1}{2}\left( p_{1}(y_{j},x_{j})h_{1}(y_{j},x_{j}) + p_{2}(y_{j},x_{j}) \right) (y_{j} - x_{j}) \nonumber \\
\label{eq:Doc36b}
	\Delta^{(r_{\uparrow})}_{3} &=& \frac{3}{2}p_{1}(x_{i},x_{j}) \Delta^{(r_{\uparrow})}_{2} . 
\eea
And similarly,
\bea
	\Delta^{r_{\downarrow}}_{2} = 2 r \left[ -(x_{i}- r ) + x_{j} \right].\\
	\Delta^{r_{\downarrow}}_{3} = \frac{3}{2} p_{1}(x_{i} , x_{j})\Delta^{r_{\downarrow}}_{2}.
\eea
We can extend this to higher $\Delta_{p}$:

\fbox{
 \addtolength{\linewidth}{-2\fboxsep}%
 \addtolength{\linewidth}{-2\fboxrule}%
 \begin{minipage}{\linewidth}
  \bea
  \Delta^{(r_{\uparrow})}_{3} &=& \frac{3}{2}p_{1}(x_{i},x_{j}) \Delta^{(r_{\uparrow})}_{2}, \\
	\Delta^{(r_{\uparrow})}_{4} &=& \frac{2}{3}p_{1}(x_{i},x_{j}) \Delta^{(r_{\uparrow})}_{3} + p_{2}(x_{i},x_{j})\Delta^{(r_{\uparrow})}_{2} + \frac{1}{2}\left(\Delta^{(r_{\uparrow})}_{2}\right)^{2},   \\
	\Delta^{(r_{\uparrow})}_{5} &=& \frac{5}{8}p_{1}(x_{i},x_{j}) \Delta^{(r_{\uparrow})}_{4} + \frac{5}{4}p_{3}(x_{i},x_{j})\Delta^{(r_{\uparrow})}_{2} + \frac{5}{8}\Delta^{(r_{\uparrow})}_{3}\,\Delta^{(r_{\uparrow})}_{2}, \\
	\Delta^{(r_{\uparrow})}_{6} &=& \frac{3}{5}p_{1}(x_{i},x_{j}) \Delta^{(r_{\uparrow})}_{5} + \frac{3}{2}p_{4}(x_{i},x_{j})\Delta^{(r_{\uparrow})}_{2} + \frac{3}{4} p_{2}(x_{i},x_{j}) \left(\Delta^{(r_{\uparrow})}_{2}\right)^{2} + \frac{1}{2}p_{1}(x_{i},x_{j})\Delta^{(r_{\uparrow})}_{3}\,\Delta^{(r_{\uparrow})}_{2} \nonumber \\
	&& + \frac{1}{4}\left(\Delta^{(r_{\uparrow})}_{2}\right)^{3},\\
	\Delta^{(r_{\uparrow})}_{7} &=& \frac{7}{6}p_{1}(x_{i},x_{j}) \Delta^{(r_{\uparrow})}_{6} + \cdots .  
  \eea
 \end{minipage}
}

\section{Generalized transitivity arguments}\label{Transitivity}

Consider figures \ref{fig1}, \ref{fig2} and \ref{fig3}. Here, we argue that the transitivity property of Young diagrams and the dominance ordering generalizes to incomparable Young diagrams for our measure of incomparability - quantifying the number of swaps of the inequality sign in the row sum relations. We only consider the case for one and two swaps, but our argument may plausibly extend to $l$ swaps. Our finding is that if $k^{(Y_{1},Y_{2})}_{swap} = l, (l = 1,2)$, and if $k^{(Y_{2},Y_{3})}_{swap} = l, (l = 1,2)$, then $k^{(Y_{1},Y_{3})}_{swap} \geq l, l = 1,2$. That is, $Y_{1}$ and $Y_{3}$ are at least as incomparable as the pairs $Y_{1} , Y_{2}$ and $Y_{2}, Y_{3}$, assuming these pairs have the same level of incomparability. To summarize, incomparability does not decrease.

\begin{center}
\begin{figure}[h!]
\begin{tikzpicture}[semithick, >=Stealth]
\coordinate (A0) at (-6,0);
\coordinate (A1) at (-2,0);
\coordinate (Abottomleft) at (-6,-4);
\draw (A0) -- (Abottomleft);
\coordinate (A2) at (-2,-0.5);
\coordinate (A3) at (-3,-0.5);
\coordinate (A4) at (-3,-1);
\coordinate (A5) at (-3.5,-1);
\coordinate (A6) at (-3.5,-1.75);
\coordinate (A7) at (-4,-1.75);
\coordinate (A8) at (-4,-2.25);
\coordinate (A9) at (-4.5,-2.25);
\coordinate (A10) at (-4.5,-3);
\coordinate (A11) at (-5,-3);
\coordinate (A12) at (-5,-3.35);
\coordinate (A13) at (-5.5,-3.35);
\coordinate (A14) at (-5.5,-4);
\coordinate (A15) at (-6,-4);
\draw (A0) -- (A1) -- (A2) -- (A3) -- (A4) -- (A5) -- (A6) -- (A7)
-- (A8) -- (A9) -- (A10) -- (A11) -- (A12) -- (A13) -- (A14) -- (A15);
\coordinate (Atop) at (-4,0);
\node[above] at (Atop) {$Y_{1}$};

\coordinate (1stMaj) at (-1,-2);
\node at (1stMaj) {\Large{$\prec$}};

\coordinate (B0) at (0,0);
\coordinate (B1) at (4,0);
\coordinate (Bbottomleft) at (0,-4);
\draw (B0) -- (Bbottomleft);
\coordinate (B2) at (4,-0.55);
\coordinate (B3) at (-3+6,-0.55);
\coordinate (B4) at (-3+6,-1.05);
\coordinate (B5) at (-3.5+6,-1.05);
\coordinate (B6) at (-3.5+6,-1.8);
\coordinate (B7) at (-4+6,-1.8);
\coordinate (B8) at (-4+6,-2.15);
\coordinate (B9) at (-4.5+6,-2.15);
\coordinate (B10) at (-4.5+6,-3);
\coordinate (B11) at (-5+6,-3);
\coordinate (B12) at (-5+6,-3.3);
\coordinate (B13) at (-5.5+6,-3.3);
\coordinate (B14) at (-5.5+6,-4);
\coordinate (B15) at (-6+6,-4);
\draw (B0) -- (B1) -- (B2) -- (B3) -- (B4) -- (B5) -- (B6) -- (B7)
-- (B8) -- (B9) -- (B10) -- (B11) -- (B12) -- (B13) -- (B14) -- (B15);
\coordinate (Btop) at (2,0);
\node[above] at (Btop) {$Y_{2}$};

\coordinate (2ndMaj) at (5,-2);
\node at (2ndMaj) {\Large{$\prec$}};

\coordinate (C0) at (6,0);
\coordinate (C1) at (10,0);
\coordinate (Cbottomleft) at (6,-4);
\draw (C0) -- (Cbottomleft);
\coordinate (C2) at (4+6,-0.65);
\coordinate (C3) at (-3+6+6,-0.65);
\coordinate (C4) at (-3+6+6,-1.0);
\coordinate (C5) at (-3.5+6+6,-1.0);
\coordinate (C6) at (-3.5+6+6,-1.65);
\coordinate (C7) at (-4+6+6,-1.65);
\coordinate (C8) at (-4+6+6,-2);
\coordinate (C9) at (-4.5+6+6,-2);
\coordinate (C10) at (-4.5+6+6,-2.7);
\coordinate (C11) at (-5+6+6,-2.7);
\coordinate (C12) at (-5+6+6,-3.5);
\coordinate (C13) at (-5.5+6+6,-3.5);
\coordinate (C14) at (-5.5+6+6,-4);
\coordinate (C15) at (-6+6+6,-4);
\draw (C0) -- (C1) -- (C2) -- (C3) -- (C4) -- (C5) -- (C6) -- (C7)
-- (C8) -- (C9) -- (C10) -- (C11) -- (C12) -- (C13) -- (C14) -- (C15);
\coordinate (Ctop) at (8,0);
\node[above] at (Ctop) {$Y_{3}$};
\end{tikzpicture}
\caption{Figure schematically showing the transitivity property for Young diagrams.}
\label{fig1}
\end{figure}
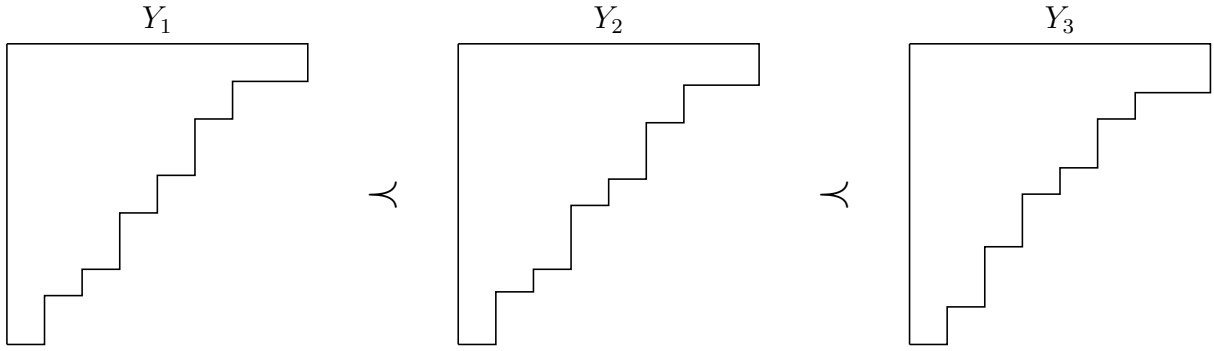
\end{center}

\begin{center}
\begin{figure}[h!]
\begin{tikzpicture}[semithick, >=Stealth]
\coordinate (A0) at (-6,0);
\coordinate (A1) at (-2,0);
\coordinate (Abottomleft) at (-6,-4);
\draw (A0) -- (Abottomleft);
\coordinate (A2) at (-2,-0.5);
\coordinate (A3) at (-3,-0.5);
\coordinate (A4) at (-3,-1);
\coordinate (A5) at (-3.5,-1);
\coordinate (A6) at (-3.5,-1.75);
\coordinate (A7) at (-4,-1.75);
\coordinate (A8) at (-4,-2.25);
\coordinate (A9) at (-4.5,-2.25);
\coordinate (A10) at (-4.5,-3);
\coordinate (A11) at (-5,-3);
\coordinate (A12) at (-5,-3.35);
\coordinate (A13) at (-5.5,-3.35);
\coordinate (A14) at (-5.5,-4);
\coordinate (A15) at (-6,-4);
\draw (A0) -- (A1) -- (A2) -- (A3) -- (A4) -- (A5) -- (A6) -- (A7)
-- (A8) -- (A9) -- (A10) -- (A11) -- (A12) -- (A13) -- (A14) -- (A15);
\coordinate (Atop) at (-4,0);
\node[above] at (Atop) {$Y_{1}$};

\coordinate (1stMajtop) at (-1,-0.8);
\node at (1stMajtop) {\Large{$\prec$}};

\coordinate (1stMajmid) at (-1,-1.95);
\node at (1stMajmid) {\Large{$\succ$}};

\coordinate (1stMajbot) at (-1,-3);
\node at (1stMajbot) {\Large{$\succ$}};

\coordinate (B0) at (0,0);
\coordinate (B1) at (4,0);
\coordinate (Bbottomleft) at (0,-4);
\draw (B0) -- (Bbottomleft);
\coordinate (B2) at (4,-0.55);
\coordinate (B3) at (-3+6,-0.55);
\coordinate (B4) at (-3+6,-1.05);
\coordinate (B5) at (-3.5+6,-1.05);
\coordinate (B6) at (-3.5+6,-1.8);
\coordinate (B7) at (-4+6,-1.8);
\coordinate (B8) at (-4+6,-2.15);
\coordinate (B9) at (-4.5+6,-2.15);
\coordinate (B10) at (-4.5+6,-3);
\coordinate (B11) at (-5+6,-3);
\coordinate (B12) at (-5+6,-3.3);
\coordinate (B13) at (-5.5+6,-3.3);
\coordinate (B14) at (-5.5+6,-4);
\coordinate (B15) at (-6+6,-4);
\draw (B0) -- (B1) -- (B2) -- (B3) -- (B4) -- (B5) -- (B6) -- (B7)
-- (B8) -- (B9) -- (B10) -- (B11) -- (B12) -- (B13) -- (B14) -- (B15);
\coordinate (Btop) at (2,0);
\node[above] at (Btop) {$Y_{2}$};

\coordinate (2ndMajtop) at (5,-0.75);
\node at (2ndMajtop) {\Large{$\prec$}};

\coordinate (2ndMajmid) at (5,-1.95);
\node at (2ndMajmid) {\Large{$\prec$}};

\coordinate (2ndMajbot) at (5,-3);
\node at (2ndMajbot) {\Large{$\succ$}};

\coordinate (C0) at (6,0);
\coordinate (C1) at (10,0);
\coordinate (Cbottomleft) at (6,-4);
\draw (C0) -- (Cbottomleft);
\coordinate (C2) at (4+6,-0.65);
\coordinate (C3) at (-3+6+6,-0.65);
\coordinate (C4) at (-3+6+6,-1.0);
\coordinate (C5) at (-3.5+6+6,-1.0);
\coordinate (C6) at (-3.5+6+6,-1.65);
\coordinate (C7) at (-4+6+6,-1.65);
\coordinate (C8) at (-4+6+6,-2);
\coordinate (C9) at (-4.5+6+6,-2);
\coordinate (C10) at (-4.5+6+6,-2.7);
\coordinate (C11) at (-5+6+6,-2.7);
\coordinate (C12) at (-5+6+6,-3.5);
\coordinate (C13) at (-5.5+6+6,-3.5);
\coordinate (C14) at (-5.5+6+6,-4);
\coordinate (C15) at (-6+6+6,-4);
\draw (C0) -- (C1) -- (C2) -- (C3) -- (C4) -- (C5) -- (C6) -- (C7)
-- (C8) -- (C9) -- (C10) -- (C11) -- (C12) -- (C13) -- (C14) -- (C15);
\coordinate (Ctop) at (8,0);
\node[above] at (Ctop) {$Y_{3}$};

\coordinate (Y1Y2left) at (-6,-1.6);
\coordinate (Y1Y2right) at (2.45,-1.6);
\draw [thick , dashed, red] (Y1Y2left) -- (Y1Y2right);
\node[left] at (Y1Y2left) {$s_{12}$};
\coordinate (Y1Y2farright) at (10,-1.6);
\draw [thick, dotted] (Y1Y2right) -- (Y1Y2farright);

\coordinate (Y2Y3left) at (0,-2.3);
\coordinate (Y2Y3right) at (10,-2.3);
\draw [thick , dashed , blue] (Y2Y3left) -- (Y2Y3right);
\coordinate (Y2Y3farleft) at (-6,-2.3);
\draw [thick, dotted] (Y2Y3left) -- (Y2Y3farleft);
\node[left] at (Y2Y3farleft) {$s_{23}$};

\end{tikzpicture}
\caption{}
\label{fig2}
\end{figure}
\end{center}

\begin{center}
\begin{figure}[h!]
\begin{tikzpicture}[semithick, >=Stealth]
\coordinate (A0) at (-6,0);
\coordinate (A1) at (-2,0);
\coordinate (Abottomleft) at (-6,-4);
\draw (A0) -- (Abottomleft);
\coordinate (A2) at (-2,-0.5);
\coordinate (A3) at (-3,-0.5);
\coordinate (A4) at (-3,-1);
\coordinate (A5) at (-3.5,-1);
\coordinate (A6) at (-3.5,-1.75);
\coordinate (A7) at (-4,-1.75);
\coordinate (A8) at (-4,-2.25);
\coordinate (A9) at (-4.5,-2.25);
\coordinate (A10) at (-4.5,-3);
\coordinate (A11) at (-5,-3);
\coordinate (A12) at (-5,-3.35);
\coordinate (A13) at (-5.5,-3.35);
\coordinate (A14) at (-5.5,-4);
\coordinate (A15) at (-6,-4);
\draw (A0) -- (A1) -- (A2) -- (A3) -- (A4) -- (A5) -- (A6) -- (A7)
-- (A8) -- (A9) -- (A10) -- (A11) -- (A12) -- (A13) -- (A14) -- (A15);
\coordinate (Atop) at (-4,0);
\node[above] at (Atop) {$Y_{1}$};

\coordinate (1stMajR1) at (-1,-0.5);
\node at (1stMajR1) {\large{$\prec$}};

\coordinate (1stMajR1still) at (-1,-1.2);
\node at (1stMajR1still) {\large{$\prec$}};

\coordinate (1stMajR2) at (-1,-2);
\node at (1stMajR2) {\large{$\succ$}};

\coordinate (1stMajR3) at (-1,-2.75);
\node at (1stMajR3) {\large{$\prec$}};

\coordinate (1stMajR3still) at (-1,-3.5);
\node at (1stMajR3still) {\large{$\prec$}};

\coordinate (B0) at (0,0);
\coordinate (B1) at (4,0);
\coordinate (Bbottomleft) at (0,-4);
\draw (B0) -- (Bbottomleft);
\coordinate (B2) at (4,-0.55);
\coordinate (B3) at (-3+6,-0.55);
\coordinate (B4) at (-3+6,-1.05);
\coordinate (B5) at (-3.5+6,-1.05);
\coordinate (B6) at (-3.5+6,-1.8);
\coordinate (B7) at (-4+6,-1.8);
\coordinate (B8) at (-4+6,-2.15);
\coordinate (B9) at (-4.5+6,-2.15);
\coordinate (B10) at (-4.5+6,-3);
\coordinate (B11) at (-5+6,-3);
\coordinate (B12) at (-5+6,-3.3);
\coordinate (B13) at (-5.5+6,-3.3);
\coordinate (B14) at (-5.5+6,-4);
\coordinate (B15) at (-6+6,-4);
\draw (B0) -- (B1) -- (B2) -- (B3) -- (B4) -- (B5) -- (B6) -- (B7)
-- (B8) -- (B9) -- (B10) -- (B11) -- (B12) -- (B13) -- (B14) -- (B15);
\coordinate (Btop) at (2,0);
\node[above] at (Btop) {$Y_{2}$};

\coordinate (2ndMajR1) at (5,-0.5);
\node at (2ndMajR1) {\large{$\prec$}};

\coordinate (2ndMajR2) at (5,-1.2);
\node at (2ndMajR2) {\large{$\succ$}};

\coordinate (2ndMajR2still) at (5,-2);
\node at (2ndMajR2still) {\large{$\succ$}};

\coordinate (2ndMajR2evenstill) at (5,-2.75);
\node at (2ndMajR2evenstill) {\large{$\succ$}};

\coordinate (2ndMajR3) at (5,-3.5);
\node at (2ndMajR3) {\large{$\prec$}};

\coordinate (C0) at (6,0);
\coordinate (C1) at (10,0);
\coordinate (Cbottomleft) at (6,-4);
\draw (C0) -- (Cbottomleft);
\coordinate (C2) at (4+6,-0.65);
\coordinate (C3) at (-3+6+6,-0.65);
\coordinate (C4) at (-3+6+6,-1.0);
\coordinate (C5) at (-3.5+6+6,-1.0);
\coordinate (C6) at (-3.5+6+6,-1.65);
\coordinate (C7) at (-4+6+6,-1.65);
\coordinate (C8) at (-4+6+6,-2);
\coordinate (C9) at (-4.5+6+6,-2);
\coordinate (C10) at (-4.5+6+6,-2.7);
\coordinate (C11) at (-5+6+6,-2.7);
\coordinate (C12) at (-5+6+6,-3.5);
\coordinate (C13) at (-5.5+6+6,-3.5);
\coordinate (C14) at (-5.5+6+6,-4);
\coordinate (C15) at (-6+6+6,-4);
\draw (C0) -- (C1) -- (C2) -- (C3) -- (C4) -- (C5) -- (C6) -- (C7)
-- (C8) -- (C9) -- (C10) -- (C11) -- (C12) -- (C13) -- (C14) -- (C15);
\coordinate (Ctop) at (8,0);
\node[above] at (Ctop) {$Y_{3}$};

\coordinate (Y1Y2left) at (-6,-1.4);
\coordinate (Y1Y2right) at (2.45,-1.4);
\draw [thick , dashed, red] (Y1Y2left) -- (Y1Y2right);
\node[left] at (Y1Y2left) {$s^{(1st)}_{12}$};
\coordinate (Y1Y2farright) at (10,-1.4);
\draw [thick, dotted] (Y1Y2right) -- (Y1Y2farright);

\coordinate (Y1Y2leftbot) at (-6,-2.5);
\coordinate (Y1Y2rightbot) at (1.45,-2.5);
\draw [thick , dashed, red] (Y1Y2leftbot) -- (Y1Y2rightbot);
\node[left] at (Y1Y2leftbot) {$s^{(2nd)}_{12}$};
\coordinate (Y1Y2farrightbot) at (10,-2.5);
\draw [thick, dotted] (Y1Y2rightbot) -- (Y1Y2farrightbot);

\coordinate (Y2Y3left) at (0,-1);
\coordinate (Y2Y3right) at (10,-1);
\draw [thick , dashed , blue] (Y2Y3left) -- (Y2Y3right);
\coordinate (Y2Y3farleft) at (-6,-1);
\draw [thick, dotted] (Y2Y3left) -- (Y2Y3farleft);
\node[right] at (Y2Y3right) {$s^{(1st)}_{23}$};

\coordinate (Y2Y3leftbot) at (0,-3);
\coordinate (Y2Y3rightbot) at (10,-3);
\draw [thick , dashed , blue] (Y2Y3leftbot) -- (Y2Y3rightbot);
\coordinate (Y2Y3farleftbot) at (-6,-3);
\draw [thick, dotted] (Y2Y3leftbot) -- (Y2Y3farleftbot);
\node[right] at (Y2Y3rightbot) {$s^{(2nd)}_{23}$};

\end{tikzpicture}
\caption{}
\label{fig3}
\end{figure}
\end{center}

Figure \ref{fig1} demonstrates that if $Y_{1}\prec Y_{2}$, and $Y_{2} \prec Y_{3}$, then $Y_{1} \prec Y_{3}$. Figure \ref{fig2} shows three diagrams for which $k^{(Y_{1},Y_{2})}_{swap} = 1$, and $k^{(Y_{2},Y_{3})}_{swap} = 1$. The swap in the row sum relations for $(Y_{1},Y_{2})$ occurs in row labeled $s_{12}$, and for $(Y_{2},Y_{3})$ in row labeled $s_{23}$. These rows are indicated by the red dashed line and the blue dashed line respectively. For $(Y_{1},Y_{2})$ the region above the red dashed line, $Y_{1} \prec Y_{2}$, and the region below the red dashed line, $Y_{1} \succ Y_{2}$. For $(Y_{2},Y_{3})$ the region above the blue dashed line, $Y_{2} \prec Y_{3}$, and the region below the red dashed line, $Y_{2} \succ Y_{3}$. With certainty, in the region above the red dashed line, $Y_{1} \prec Y_{3}$, and in the region below the blue dashed line, $Y_{1} \succ Y_{3}$. However in the region between the two dashed lines, it is not certain which region dominates. However, we may conclude then is that there is a minimum of one swap in the row sum relations for $Y_{1}$ and $Y_{3}$. This swap occurs somewhere between (and possibly including) the two dashed lines. 

 Figure \ref{fig3} shows three diagrams for which $k^{(Y_{1},Y_{2})}_{swap} = 2$, and $k^{(Y_{2},Y_{3})}_{swap} = 2$. The first swap in the row sum relations for $(Y_{1},Y_{2})$ occurs in row labeled $s^{(1st)}_{12}$, and for $(Y_{2},Y_{3})$ in row labeled $s^{(1st)}_{23}$. These rows are again indicated by the top red dashed line and the top blue dashed line respectively. The second swap in the row sum relations for $(Y_{1},Y_{2})$ occurs in row labeled $s^{(2nd)}_{12}$ (bottom red dashed line), and for $(Y_{2},Y_{3})$ in row labeled $s^{(2nd)}_{23}$ (blue dashed line). In the region above the first blue dashed line, $Y_{1}\prec Y_{2} \prec Y_{3}$. In the region between the red dashed lines, $Y_{1} \succ Y_{2} \succ Y_{3}$, and in the region below the bottom blue dashed line, we have $Y_{1}\prec Y_{2} \prec Y_{3}$. We cannot compare the regions of $(Y_{1} , Y_{3})$ between the top blue and red dashed lines, and similarly, we cannot compare the regions of $(Y_{1} , Y_{3})$ between the bottom blue and red dashed lines. However, we may conclude that $Y_{1}$ and $Y_{3}$ have at least two swaps in their row sum relations. The first swap occurs somewhere between (and possibly including) the top dashed lines, and the second swap occurs somewhere between (and possibly including) the bottom dashed lines.

\section{Building partitions with $n^{(l)}_{min}$ boxes }\label{ConstructPart}

\subsection{One swap}

We write the partitions in this section in non-decreasing order. Begin the construction of $Y$ and $Y'$ with the first entry of $Y$ being 0, and the first entry of $Y'$ being 1. Thus, we begin with $Y \prec Y'$. Next, choose the second entry of $Y$ to be 2, and the second entry of $Y'$ to be 1 again. So far 
\bea
	Y &=& (0,2)\\
	Y' &=& (1,1),
\eea
with the symbol for $(Y,Y')$ being $(<,=)$. Now, let us generate the first swap with
\bea
	Y &=& (0,2,2)\\
	Y' &=& (1,1,1).
\eea
The symbol for $(Y,Y')$ is now $(<,=,>)$. We want both $Y$ and $Y'$ to be partitions of the same $n$. Attach the smallest possible integers to the end of $Y$ and $Y'$ so that they have the same $n$ - these are a $2$ to the end of $Y$ and a 3 to the end of $Y'$. This gives
\bea
\label{eq:7a}
	Y &=& (0,2,2,2)\\
\label{eq:7b}
	Y' &=& (1,1,1,3).
\eea
The symbol is $(<,=,>,=)$ and adding the parts of (\ref{eq:7a}), or (\ref{eq:7b}), we see $n^{(l=1)}_{min}=6$. Notice that besides the $=$ at the end which is mandatory there is only one equality sign in the row sum relations. There also exists another pair of Young diagrams with six boxes but with no equality in the row sums until the very end. This pair is
\bea
\label{eq:7a1}
	Y &=& (0,3,3),\\
\label{eq:7b1}
	Y' &=& (1,1,4).
\eea
Here, the symbol is $(<,>,=)$.

\subsection{Two swaps}

For two swaps, we begin in the same way as above with the entries: 
\bea
	Y &=& (0,2,2),\\
	Y' &=& (1,1,1).
\eea
So far, $Y\vdash 4$, $Y' \vdash 3$, and we have generated one swap. We now generate the second swap by adding the smallest integers to $Y$ and $Y'$ so that they are both valid diagrams and produce a swap. Thus,
\bea
\label{eq:7c}
	Y &=& (0,2,2,2),\\
\label{eq:7d}
	Y' &=& (1,1,1,4).
\eea
The symbol for $(Y,Y')$ is $(<,=,>,<)$ but $Y \vdash 6$ and $Y'\vdash 7$. The last integer should result in both $Y$ and $Y'$ being partitions of $n$. The smallest possible entries are $4$ for $Y'$ and $5$ for $Y$. Thus,
\bea
\label{eq:7e}
	Y &=& (0,2,2,2,5),\\
\label{eq:7f}
	Y' &=& (1,1,1,4,4),
\eea
whose symbol is $(<,=,>,<,=)$. Now both $Y$ and $Y'$ are partitions of 11. From this construction, $n^{(l=2)}_{min} = 11$. 

\subsection{Three swaps}

Before stating the general pattern we followed, we provide one more explicit construction. We construct a pair having three swaps. Like the $l=1$ case, we will see that we can construct two pairs. For the first pair, begin from
\bea
\label{eq:7g}
	Y &=& (0,2,2,2),\\
\label{eq:7h}
	Y' &=& (1,1,1,4),
\eea
whose symbol is $(<,=,>,<,=)$ - we have already generated two swaps and one $=$ sign. We generate the next swap immediately by extending $Y$ and $Y'$ using the smallest possible integers to obtain
\bea
\label{eq:7g}
	Y &=& (0,2,2,2,6),\\
\label{eq:7h}
	Y' &=& (1,1,1,4,4).
\eea
The symbol so far is $(<,=,>,<,>)$ and $Y$ and $Y'$ have three swaps. Lastly, extend $Y$ and $Y'$ using the smallest integers possible to maintain valid Young diagrams and have the same $n$ - they are 6 for $Y$ and $7$ for $Y'$. We now have
\bea
\label{eq:7i}
	Y &=& (0,2,2,2,6,6),\\
\label{eq:7j}
	Y' &=& (1,1,1,4,4,7).
\eea
The final symbol for this $(Y,Y')$ is $(<,=,>,<,>,=)$. Our construction yields $n^{(l=3)}_{min}$. Note only 1 $=$ sign in the row sum relations (apart from the one appearing at the end). We can also construct a pair of Young diagrams with 18 boxes also having three swaps but with two $=$ signs in their row sum relations. The pair is
\bea
\label{eq:7k}
	Y &=& (0,2,2,2,2,5,5),\\
\label{eq:7l}
	Y' &=& (1,1,1,3,3,3,6).
\eea
The symbol here is $(<,=,>,=,<,>,=)$.

For $l=4$,
\bea
\label{eq:Da}
	Y &=& (0,2,2,2,2,5,5,8), \nonumber \\
	Y' &=& (1,1,1,3,3,3,7,7).
\eea

For $l=5$,
\bea
\label{eq:Db}
	Y_{1} &=& (0,2,2,2,2,4,4,4,8,8),\nonumber\\
	Y'_{1} &=& (1,1,1,3,3,3,3,6,6,9).\\
\label{eq:Dc}
	Y_{2} &=& (0,2,2,2,2,5,5,9,9),\nonumber\\
	Y'_{2} &=& (1,1,1,3,3,3,7,7,10).
\eea

For $l=6$,
\bea
\label{eq:Dd}
	Y_{1} &=& (0,2,2,2,2,4,4,4,8,8,11),\nonumber\\
	Y'_{1} &=& (1,1,1,3,3,3,3,6,6,10,10).
\eea

For $l=7$,
\bea
\label{eq:Dd}
	Y_{1} &=& (0, 2, 2, 2, 2, 4, 4, 4, 8, 8, 12,12),\nonumber\\
	Y'_{1} &=& (1, 1, 1, 3, 3, 3, 3, 6, 6 , 10 , 10 , 13).\\
	Y_{2} &=& (0, 2, 2, 2, 2, 4, 4, 4, 4, 7, 7, 11, 11) \nonumber \\
	Y'_{2} &=& (1, 1, 1, 3, 3, 3, 3, 5, 5, 5, 9, 9, 12).
\eea

For $l=8$,
\bea
\label{eq:Dd}
	Y_{1} &=& (0, 2, 2, 2, 2, 4, 4, 4, 4, 7, 7, 11, 11, 14),\nonumber\\
	Y'_{1} &=& (1, 1, 1, 3, 3, 3, 3, 5, 5, 5, 9, 9, 13, 13).
\eea

For $l=9$,
\bea
\label{eq:Dd}
	Y_{1} &=& (0, 2, 2, 2, 2, 4, 4, 4, 4, 7, 7, 11, 11, 15, 15),\nonumber\\
	Y'_{1} &=& (1, 1, 1, 3, 3, 3, 3, 5, 5, 5, 9, 9, 13, 13, 16).\\
	Y_{2} &=& (0, 2, 2, 2, 2, 4, 4, 4, 4, 6, 6, 6, 10, 10, 14, 14) \nonumber \\
	Y'_{2} &=& (1, 1, 1, 3, 3, 3, 3, 5, 5, 5, 5, 8, 8, 12, 12, 15).
\eea

For $l=10$,
\bea
\label{eq:Dd}
	Y_{1} &=& (0, 2, 2, 2, 2, 4, 4, 4, 4, 6, 6, 6, 10, 10, 14, 14, 17),\nonumber\\
	Y'_{1} &=& (1, 1, 1, 3, 3, 3, 3, 5, 5, 5, 5, 8, 8, 12, 12, 16, 16).
\eea

For $l=11$,
\bea
\label{eq:Db}
	Y_{1} &=& (0,2,2,2,2,4,4,4,4,6,6,6,6,9,9,13,13,17,17),\nonumber\\
	Y'_{1} &=& (1,1,1,3,3,3,3,5,5,5,5,7,7,7,11,11,15,15,18).\\
\label{eq:Dc}
	Y_{2} &=& (0,2,2,2,2,4,4,4,4,6,6,6,10,10,14,14,18,18),\nonumber\\
	Y'_{2} &=& (1,1,1,3,3,3,3,5,5,5,5,8,8,12,12,16,16,19).
\eea

\end{document}